\documentclass[twocolumn]{aastex63}

\usepackage{amssymb, amsfonts, amsmath,framed}
\usepackage{comment}
\usepackage{enumitem}

\usepackage{bm}
\expandafter\ifx\csname package@font\endcsname\relax\else
 \expandafter\expandafter
 \expandafter\usepackage
 \expandafter\expandafter
 \expandafter{\csname package@font\endcsname}
\fi
\expandafter\ifx\csname package@font\endcsname\relax\else
 \expandafter\expandafter
 \expandafter\usepackage
 \expandafter\expandafter
 \expandafter{\csname package@font\endcsname}
\fi
\def\bi{\begin{itemize}}
\def\ei{\end{itemize}}
\def\bq{\begin{equation}}
\def\eq{\end{equation}}
\def\bqy{\begin{eqnarray}}
\def\eqy{\end{eqnarray}}

\begin{document}
\title{Concurrent Particle Acceleration and Pitch-Angle Anisotropy Driven by Magnetic Reconnection: Ion-Electron Plasmas} 

\correspondingauthor{}
\email{luca.comisso@columbia.edu}

\author{Luca Comisso}
\affiliation{Department of Astronomy and Columbia Astrophysics Laboratory, Columbia University, New York, New York 10027, USA}

\begin{abstract}

Particle acceleration and pitch-angle anisotropy resulting from magnetic reconnection are investigated in highly magnetized ion-electron plasmas. 
By means of fully kinetic particle-in-cell simulations, we demonstrate that magnetic reconnection generates anisotropic particle distributions $f_s \left( {|\cos \alpha|,\varepsilon} \right)$, characterized by broken power laws in both the particle energy spectrum $f_s (\varepsilon) \propto \varepsilon^{-p}$ and the pitch angle $\langle \sin^2 \alpha \rangle \propto \varepsilon^{m}$.
The characteristics of these distributions are determined by the relative strengths of the magnetic field's guide and reconnecting components ($B_g/B_0$) and the plasma magnetization ($\sigma_0$). 
Below the injection break energy $\varepsilon_0$, ion and electron energy spectra are extremely hard ($p_< \lesssim 1$) for any $B_g/B_0$ and $\sigma_0 \gtrsim 1$, while above $\varepsilon_0$, the spectral index steepens ($p_> \gtrsim 2$), displaying high sensitivity to both $B_g/B_0$ and $\sigma_0$. 
The pitch angle displays power-law ranges with negative slopes ($m_<$) below and positive slopes ($m_>$) above $\varepsilon_{\min \alpha}$, steepening with increasing $B_g/B_0$ and $\sigma_0$.
The ratio $B_g/B_0$ regulates the redistribution of magnetic energy between ions ($\Delta{E_i}$) and electrons ($\Delta{E_e}$), with $\Delta{E_i} \gg \Delta{E_e}$ for $B_g/B_0 \ll 1$, $\Delta{E_i} \sim \Delta{E_e}$ for $B_g/B_0 \sim 1$, and $\Delta{E_i} \ll \Delta{E_e}$ for $B_g/B_0 \gg 1$, with $\Delta{E_i}/\Delta{E_e}$ approaching unity when $\sigma_0 \gg 1$. 
The anisotropic distribution of accelerated particles results in an optically thin synchrotron power spectrum $F_\nu(\nu) \propto\nu^{(2-2p+m)/(4+m)}$ and a linear polarization degree $\Pi_{\rm lin} = (p+1)/(p+7/3+m/3)$ for a uniform magnetic field. Pitch-angle anisotropy also induces temperature anisotropy and eases synchrotron cooling, along with producing beamed radiation aligned with the magnetic field, potentially responsible for rapid frequency-dependent variability.

\vspace{0.9cm}

\end{abstract}

\section{Introduction} \label{sec:intro} 

Magnetic reconnection stands out as an efficient and rapid physical process to release magnetic field energy in astrophysical environments  \citep{PF2000,Kulsrud2005,Ji2022Nat}. 
It is believed to play a central role in a variety of astrophysical systems, including 
solar and stellar coronae \citep[e.g.][]{ShiMag11,Zimovets21},  
pulsar wind nebulae \citep[e.g.][]{LK2001,Cerutti13,Lyut18}, 
accretion flows around black holes \citep[e.g.][]{deGouveia05,Ripperda22}, and jetted outflows from compact objects \citep[e.g.][]{DrenkhahnSpruit02,Gia09,Sob23}. 
In these systems, the dissipation of magnetic fields via magnetic reconnection is often invoked to explain the generation of nonthermal particles and their associated radiative signatures \citep[e.g.][]{Zenitani01,Jaroschek04,Lyub08,Cerutti12,Sironi14,Guo14,Yuan16}. 

When attempting to decipher the radiative emission from the energized particles, accounting for their energy distribution alone may not be sufficient to determine the emission's properties. In environments with significant magnetic fields, such as when magnetic reconnection is at play, knowledge of the pitch angle distribution becomes necessary \citep[e.g.][]{RybLig79,Shu91,Longair2011}. 
It is customary to assume that the emitting particles have an isotropic distribution in pitch angle, based on the premise that the underlying acceleration mechanism does not imprint strong anisotropies, or that some other process (e.g., plasma instabilities) is capable of isotropizing the distribution on timescales shorter than the particle cooling time \citep[e.g.][]{Hasegawa75,Melrose86,Gary93}. 
However, in magnetically dominated collisionless plasmas, magnetic reconnection—whether occurring in turbulent environments \citep{Comisso18,Comisso19,Comisso20ApJL,Comisso21,Comisso22} or in isolation \citep{Comisso23}—can produce highly anisotropic pitch angle distributions. 
Moreover, in highly magnetized (low-$\beta$) plasmas, kinetic instabilities such as firehose and mirror modes do not constrain the resulting pitch angle anisotropy.

Accurately characterizing the energy and pitch-angle distributions of particles energized by reconnection is essential for interpreting synchrotron emission and developing reliable radiative models  \citep[e.g.][]{Lloyd2000,YangZhang18,Sobacchi_SSC2021,Goto22,Galishnikova23}. 
Generally, pitch-angle anisotropy varies with particle energy \citep{Comisso19,Comisso22}, influencing the synchrotron spectrum and radiative cooling \citep{Comisso20ApJL,Comisso21}. 
A characterization of its energy dependence was provided in \citet{Comisso23} for pair plasmas. It was found that reconnection produces broken power laws in both the electron/positron energy spectrum and pitch-angle anisotropy. 
It was also shown that these broken power laws are governed by the relative strength of the guide field compared to the reconnecting component of the magnetic field, along with the electron-positron magnetization.
However, a significant gap remains in our understanding of the energy and pitch-angle distributions resulting from magnetic reconnection in ion-electron plasmas. Addressing this gap is essential for determining the outcomes of magnetic reconnection in ion-electron plasmas, such as those found in stellar coronae \citep{HubenyMihalas15}, hot accretion flows around black holes \citep{FKRbook2002}, and jets from active galactic nuclei \citep{Krolik99}.

In this paper, we expand on the work of \citet{Comisso23} by investigating the simultaneous generation of energetic particles and pitch-angle anisotropy in ion-electron plasmas. 
We demonstrate that particle acceleration by magnetic reconnection produces broken power laws in both the energy spectrum and pitch-angle anisotropy, for both ions and electrons. 
We assess how the power-law slopes and break energies depend on the guide field strength and plasma magnetization. 
We also evaluate how the strength of the guide field affects the reconnection rate and governs the redistribution of the converted magnetic energy into ions and electrons.
Lastly, we examine the implications of the obtained particle energy and pitch-angle distributions in regulating synchrotron luminosity, spectral energy distribution, polarization of synchrotron radiation, particle cooling, the radiation-reaction limit, emission beaming, and temperature anisotropy.

\section{Numerical Method and Setup} \label{sec:method} 
We solve the relativistic Vlasov-Maxwell system of equations with the Particle-in-Cell (PIC) method \citep{birdsall_langdon_85} using the massively parallel code TRISTAN-MP \citep{buneman_93, spitkovsky_05}. 
We adopt a two-dimensional spatial domain where we evolve all three components of the electromagnetic field and particle momenta. 
The computational domain is periodic in the $x$-direction and continuously expands in the $y$-direction, where two injectors, moving away from the reconnection layer at the speed of light, introduce fresh magnetized plasma into the simulation domain (see \citet{Sironi14} for additional details).

The plasma consists of electrons and ions, with an ambient particle density of $n_{e0} + n_{i0} = n_0$. Computational particles are initialized following a Maxwell-J\"{u}ttner distribution with temperature $T_{e0} = T_{i0} = T_0$, where $T_{e0}$ and $T_{i0}$ are the electron and ion temperatures, respectively.
We adopt a Harris sheet equilibrium \citep{Harris62} with magnetic field given by $\bm{B} = B_0 \tanh{(y/\lambda)}\bm{\hat{x}}+B_g\bm{\hat{z}}$, where $B_0$ denotes the reconnecting magnetic field component, $B_g$ indicates the guide field component, and $\lambda$ is the current sheet half-thickness. The particle density profile across the current sheet follows $n = n_0 [1 + 3 \cosh^{-2}(y/\lambda)]$. Electrons within the current sheet have a drift velocity in the $z$-direction to satisfy Amp\'ere's law.

The tearing stability index \citep{FKR63} of the initial current sheet is $\Delta' = (2/\lambda) [(k \lambda)^{-1} - k\lambda]$, where $k$ indicates the perturbation wavenumber. The current sheet is unstable to modes with $k\lambda < 1$. We select $\lambda = d_i$, where $d_i=c/\omega_{pi}=c/(4 \pi n_{i0} e^2/m_i)^{1/2}$ is the ion inertial length, with $m_i$ denoting the ion mass and $e$ indicating the elementary charge (a charge number $Z=1$ is assumed). We perturb the Harris sheet equilibrium with a small-amplitude, long-wavelength perturbation. We set the length of the domain along the current sheet direction to $2 L_x = 300 d_i$, where $L_x$ denotes the half-length of the current sheet. We discretize the domain using a grid with cell size $\Delta x = \Delta y = 0.25 d_e$, where $d_e=c/\omega_{pe}=c/(4 \pi n_{e0} e^2/m_e)^{1/2}$ is the electron inertial length, and employ 20 particles per cell for the ambient plasma. We use a reduced ion-to-electron mass ratio of $m_i/m_e = 100$ to make the simulations tractable.

We conduct a campaign of 35 simulations characterized by varying ratios of the guide to reconnecting magnetic field ($B_g/B_0$) and different values of the magnetization parameter ($\sigma_0$). Here, the magnetization parameter 
\begin{equation}
   \sigma_0 = \frac{B_0^2}{4\pi (n_{e0} m_e + n_{i0} m_i) c^2} 
\end{equation} 
is the plasma magnetization associated with the reconnecting magnetic field. The plasma magnetization associated with the total magnetic field is $\sigma = \sigma_0 + \sigma_g$, where $\sigma_g = {B_g^2}/{[4\pi c^2(n_{e0} m_e + n_{i0} m_i)]}$. 
We scan 7 values of the guide to reconnecting field ratio, $B_g/B_0 = (1/8, 1/4, 1/2, 1, 2, 4, 8)$, and 5 different values of plasma magnetization, $\sigma_0 = (1/4, 1/2, 1, 2, 4)$. We consider a temperature of $k_B T_0 = 0.1 m_e c^2$ for the ambient plasma. The plasma $\beta$ associated with the reconnecting magnetic field is $\beta_0 = {4 k_B T_0}/{\sigma_{0,i} m_i c^2} \simeq (1.6, 0.8, 0.4, 0.2, 0.1) \times 10^{-2}$, where $\sigma_{0,i} = {B_0^2}/{4\pi n_{i0} m_i c^2} = \sigma_0 (1+ m_e/m_i)$.
We set the numerical speed of light to 0.45 cells per time step and run all simulations for a minimum duration of $t = 5 (L_x/v_{A0}) \max [1, B_g/B_0]$, where $v_{A0} = c [\sigma_0/(1+\sigma_0)]^{1/2}$ is the Alfv{\'e}n speed associated with the reconnecting field. This choice ensures that we have an adequate time range to assess the steady-state properties of the system.

\section{Numerical Results} \label{sec:results} 

\begin{figure}
\begin{center}
    \includegraphics[width=8.65cm]{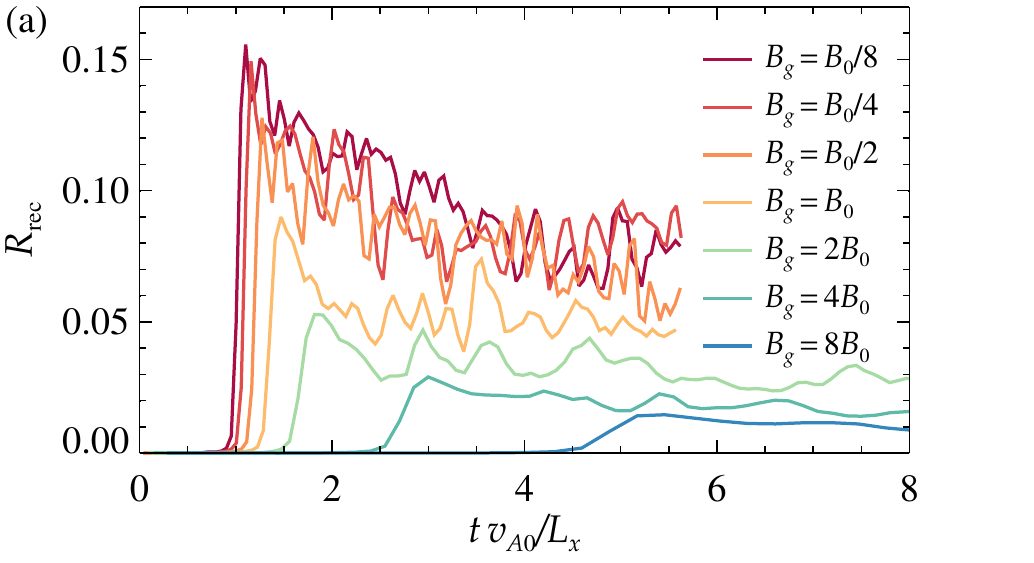}

\vspace{0.3cm}
    
    \includegraphics[width=8.65cm]{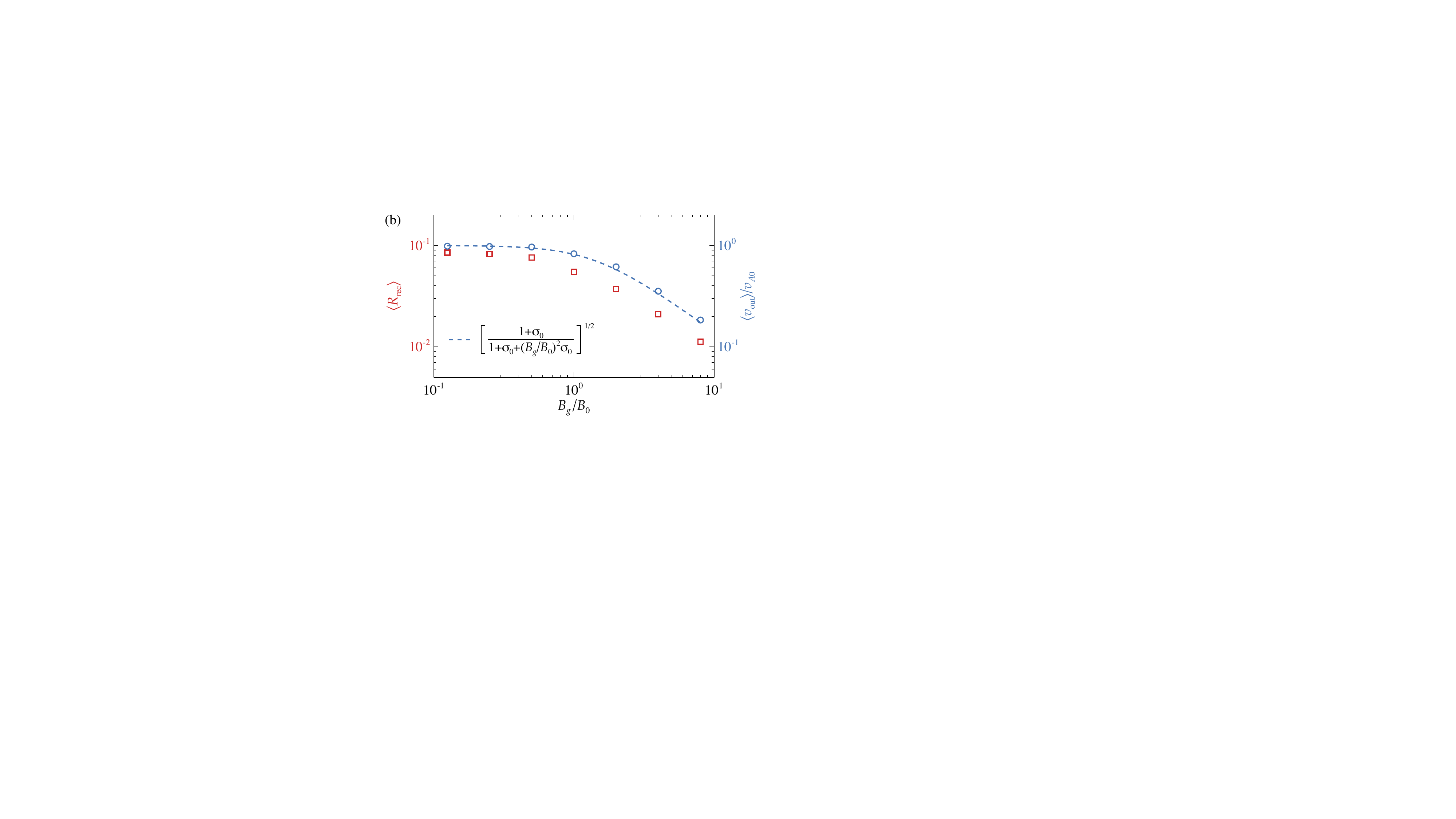}

\vspace{0.3cm}
    
    \includegraphics[width=8.65cm]{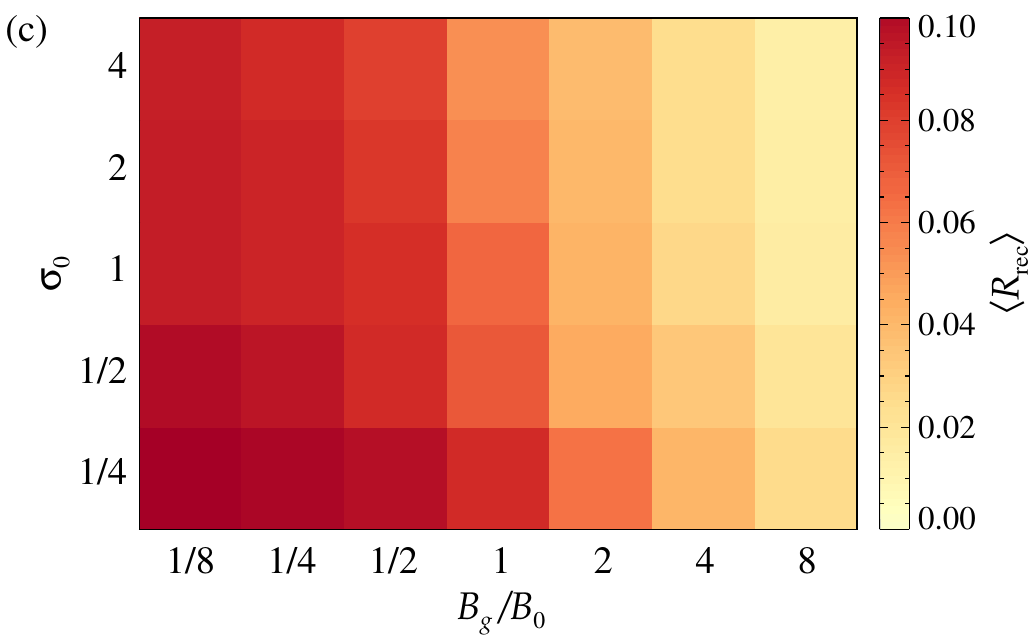}
\end{center}
\vspace{-0.2cm}
    \caption{(a) Time evolution of the reconnection rate $R_{\rm rec}$ for simulations with $B_g = (1/8, 1/4, 1/2, 1, 2, 4, 8) B_0$ and $\sigma_0 = 1$. (b) Time-averaged reconnection rate, $\langle R_{\rm rec} \rangle$, and time-averaged maximum outflow velocity, $\langle v_{\rm out} \rangle$ (normalized by $v_{A0}$), during the statistical steady-state regime ($t \gtrsim L_x/v_{A0}$ after the peak of the reconnection rate) for the same simulations. A dashed blue line indicates $[(1+\sigma_0)/(1+\sigma_0+(B_g/B_0)^2 \sigma_0)]^{1/2}$.  
(c) 2D histogram showing the variation of $\langle R_{\rm rec} \rangle$ across the parameter space of $B_g/B_0$ and $\sigma_0$.}
\label{fig1}
\end{figure}

We show the time evolution of the reconnection rate for simulations with varying guide field strength in Figure \ref{fig1}(a). The reconnection rate, normalized with respect to $v_{A0} B_0$, is computed as
\begin{equation}
   R_{\rm rec} = \frac{c E_{\rm rec}}{v_{A0} B_0} = \frac{1}{v_{A0} B_0} \frac{\partial}{\partial t} \Big(\max(A_z) - \min(A_z)\Big) \, ,
\end{equation}
where $E_{\rm rec}$ denotes the reconnection electric field and $A_z$ indicates the $z$-component of the magnetic vector potential, evaluated along $y=0$. 
With increasing guide field strength, the onset of fast reconnection is delayed \citep{PC2004}. After the onset, the reconnection rate approaches a statistical steady state, characterized by a rate that decreases with higher guide field strength. 

\begin{figure*}
    \centering
     \includegraphics[width=18.4cm]{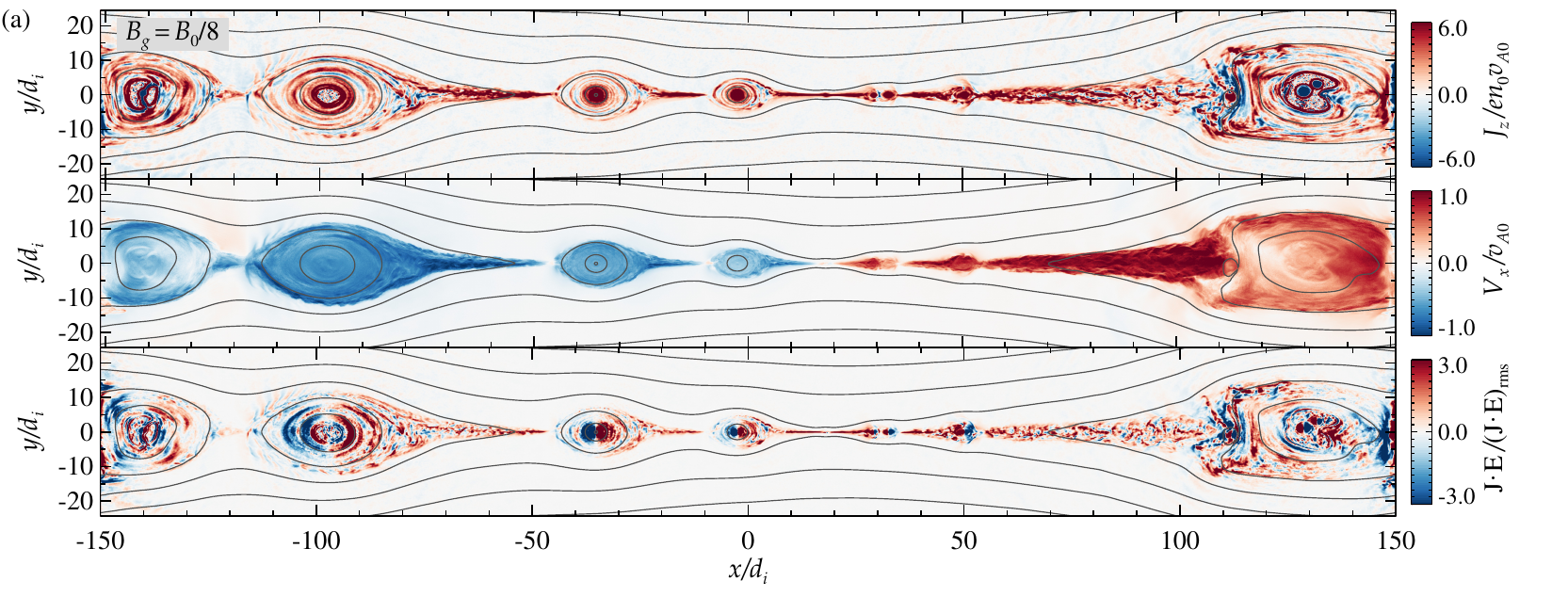}
     \includegraphics[width=18.4cm]{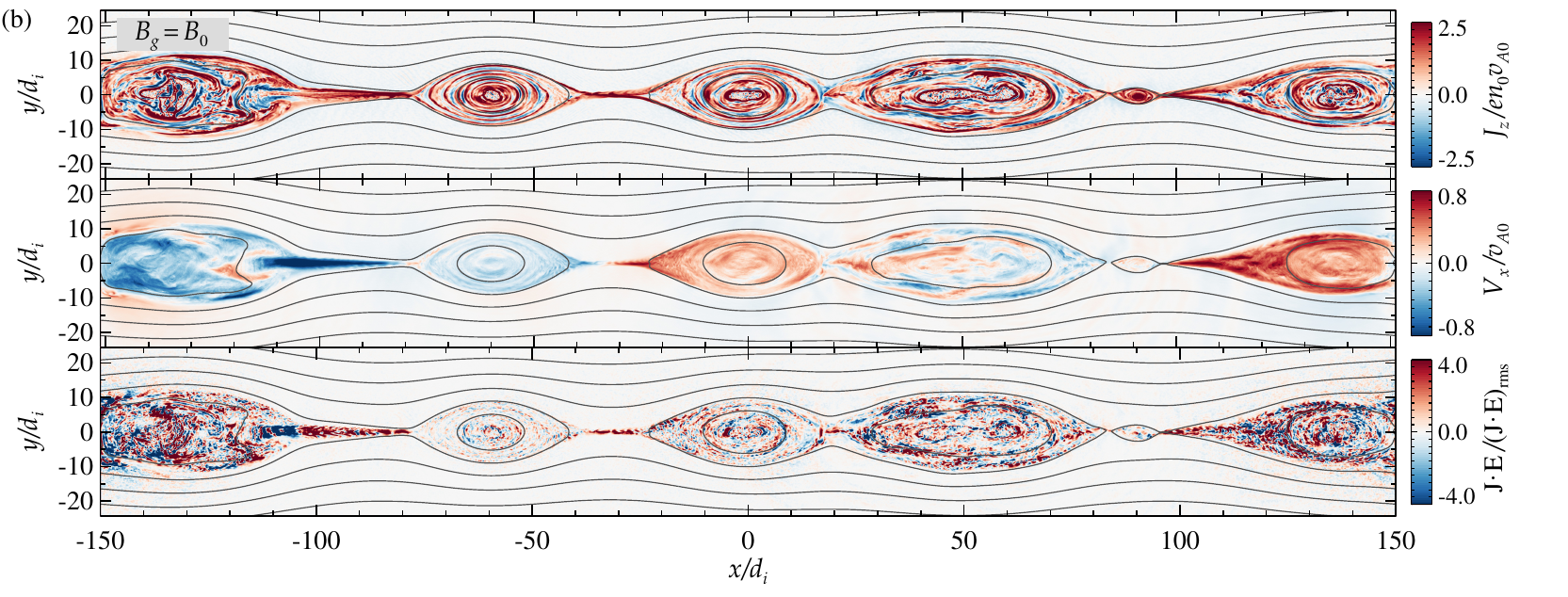}
     \includegraphics[width=18.4cm]{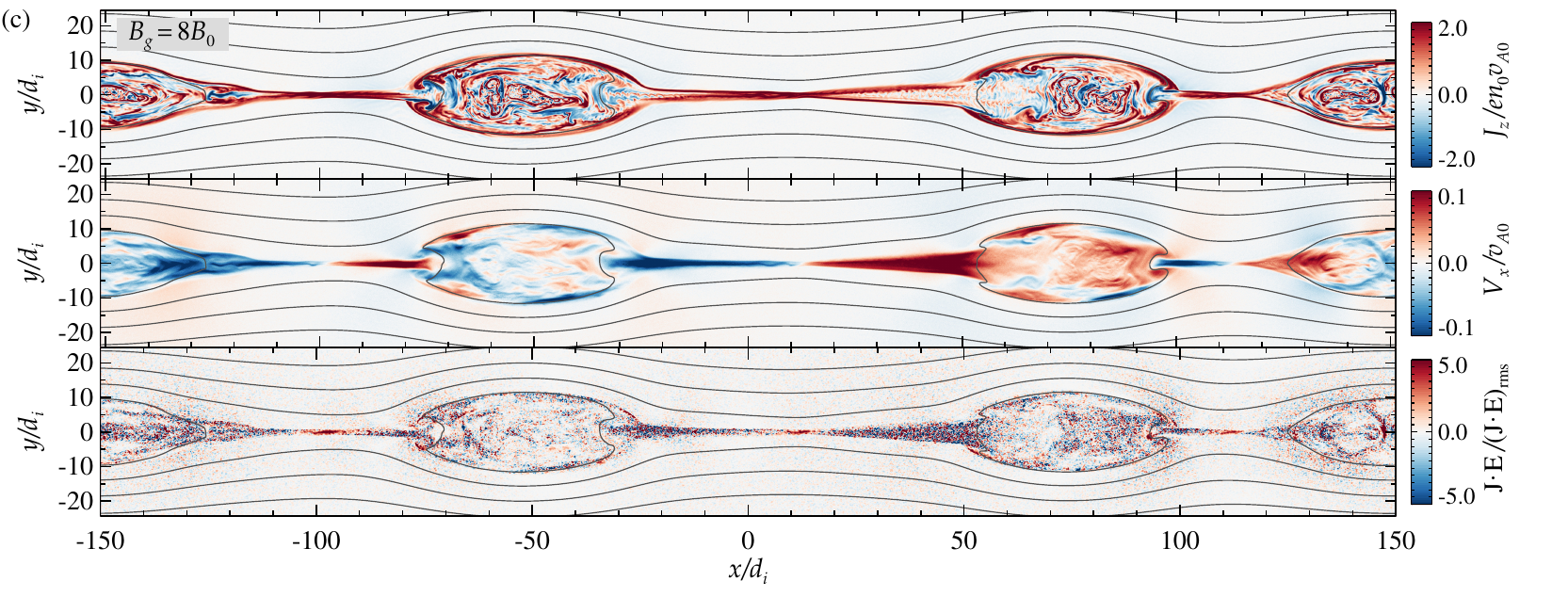}
    \caption{Visualization of different quantities within the magnetic reconnection layer during the statistical steady-state regime, for simulations with varying guide field strengths: (a) $B_g=B_0/8$, (b) $B_g=B_0$, and (c) $B_g=8B_0$, alongside magnetization $\sigma_0=1$. The panels illustrate the out-of-plane current density $J_z$ (normalized by $e n_0 v_{A0}$), the plasma velocity along the outflow direction (normalized by $v_{A0}$), and the energy conversion rate through the work done by the electric field on the particles, $\bm{J} \cdot \bm{E}$ (normalized by its root-mean-square value). Snapshots are taken at times approximately $t \sim 2 L_x/\langle v_{\rm out} \rangle$, capturing similar stages of nonlinear evolution. The displayed domain region is restricted to $|y| \leq 24 d_i$ in order to emphasize the small-scale structures in the reconnection layer.}
    \label{fig2}        
\end{figure*} 

In Figure \ref{fig1}(b), we show the time-averaged reconnection rate, $\langle R_{\rm rec} \rangle$, alongside the time-averaged maximum outflow velocity, $\langle v_{\rm out} \rangle$, during the statistical steady-state regime. Both $\langle R_{\rm rec} \rangle$ and $\langle v_{\rm out} \rangle$ decrease with increasing guide field strength. The outflow velocity is constrained by the in-plane Alfv\'en speed \citep{PF2000}, which is given by the projection of the total Alfv\'en speed into the reconnection plane, $v_A \cos \zeta$, where $v_A= c [(\sigma_0+\sigma_g)/(1+\sigma_0+\sigma_g)]^{1/2}$ and $\zeta = \arctan (B_g/B_0)$. Therefore, the maximum outflow velocity can be expressed as
\begin{equation} \label{eq:vout} 
\langle v_{\rm out} \rangle \simeq c \left( {\frac{\sigma_0}{1 + \sigma_0 + (B_g/B_0)^2 \sigma_0}} \right)^{1/2} \, .
\end{equation}
For $\sigma_0 \gg 1$, $\langle v_{\rm out} \rangle \simeq c B_0/{({B_0^2 + B_g^2})^{1/2}}$. 
The reconnection rate closely tracks $\langle v_{\rm out} \rangle/v_{A0}$, as the reconnection electric field scales with the outflow velocity in an approximate steady state, 
\begin{equation} 
\langle E_{\rm rec} \rangle \simeq B_0 \langle v_{\rm in}/c \rangle \simeq \epsilon_{\rm rec} B_0 \langle v_{\rm out}/c \rangle \, .
\end{equation} 
The factor $\epsilon_{\rm rec}$ is approximately $\mathcal{O}(10^{-1})$ for fast collisionless reconnection \citep{ComissoJPP16,CassakJPP17}. 
Therefore, an expression for the reconnection rate applicable for arbitrary guide field magnitudes is given by 
\begin{equation} \label{eq:Rrec} 
\langle R_{\rm rec} \rangle \simeq \epsilon_{\rm rec} \left( {\frac{1+\sigma_0}{1 + \sigma_0 + (B_g/B_0)^2 \sigma_0}} \right)^{1/2} \, , 
\end{equation}
with the relevant limits
\begin{equation} 
\langle R_{\rm rec} \rangle \simeq \epsilon_{\rm rec} \, , \qquad \frac{B_g}{B_0} \ll \frac{c}{v_{A0}} \, ,
\end{equation}
\begin{equation} 
\langle R_{\rm rec} \rangle \simeq \epsilon_{\rm rec} \frac{c B_0}{v_{A0} B_g} \, , \qquad \frac{B_g}{B_0} \gg \frac{c}{v_{A0}} \, .
\end{equation}
In Figure \ref{fig1}(b), we show that Eq. (\ref{eq:Rrec}), with $\epsilon_{\rm rec} \sim 0.1$, accurately describes the scaling of the reconnection rate across the range of guide field magnitudes explored. 
Additionally, in Fig. \ref{fig1}(c), we present a comprehensive summary of the steady-state reconnection rate measured throughout our simulation campaign, highlighting the decrease in the reconnection rate with increasing $B_g/B_0$, irrespective of whether $\sigma_0$ exceeds or falls below $1$ (i.e., regardless of whether the regime is relativistic or not).

In Figure \ref{fig2}, we present representative snapshots of the reconnection layer for $B_g = (1/8, 1, 8) B_0$ and $\sigma_0=1$ at comparable stages of nonlinear evolution. The fragmentation of the reconnection layer leads to the formation of plasmoids \citep{Comisso2016PoP,Comisso2017ApJ,Uzd2016}—distinct island-like structures formed within the reconnecting current sheet.
As $B_g/B_0$ increases, the out-of-plane current density concentrates more prominently along the reconnection separatrices (compare the top frames of Figs. \ref{fig2}(a) and \ref{fig2}(c)), resulting in the formation of an open Petschek-like configuration when $B_g \gg B_0$ \citep{Comisso2013}, which effectively inhibits plasmoid formation in the presence of a strong guide field. 
Furthermore, as $B_g/B_0$ increases, the reduction in the outflow velocity becomes significant, with an order of magnitude decrease from $B_g = B_0/8$ to $B_g = 8 B_0$ (compare the color bar values in the middle frames of Figs. \ref{fig2}(a) and \ref{fig2}(c)). 
Lastly, in the bottom frames of Figs. \ref{fig2}(a)-(c), the plasma energization rate $\bm{J} \cdot \bm{E}$ illustrates that under low guide field strengths, most plasma energization takes place within outflows and plasmoids \citep{Drake06,XLi15,Arnold21}. Conversely, with higher guide field strength, regions of plasma energization primarily localize around $X$-points and separatrices \citep{Pritchett06,Wang2016ApJ,McCubbin22}.

\begin{figure*}
 \centering 
  \includegraphics[width=8.65cm]{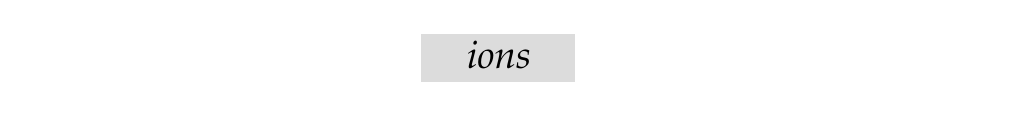}  \hspace{0.3cm}
  \includegraphics[width=8.65cm]{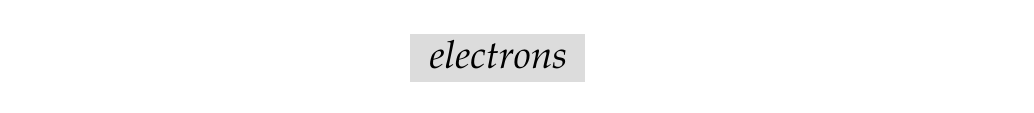}
  \hfill
  \includegraphics[width=8.65cm]{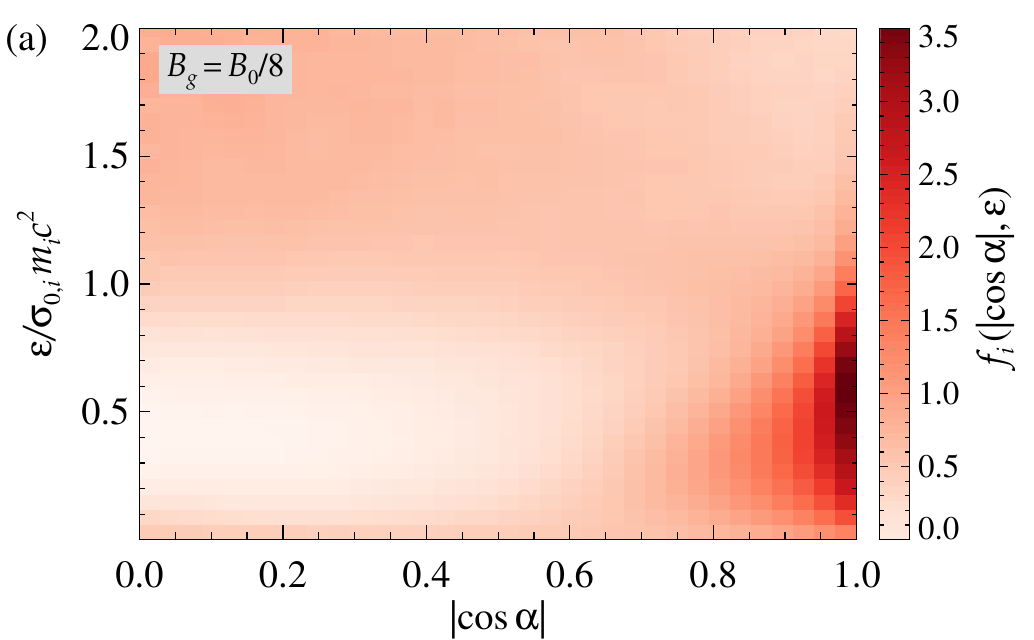} \hspace{0.3cm} 
  \includegraphics[width=8.65cm]{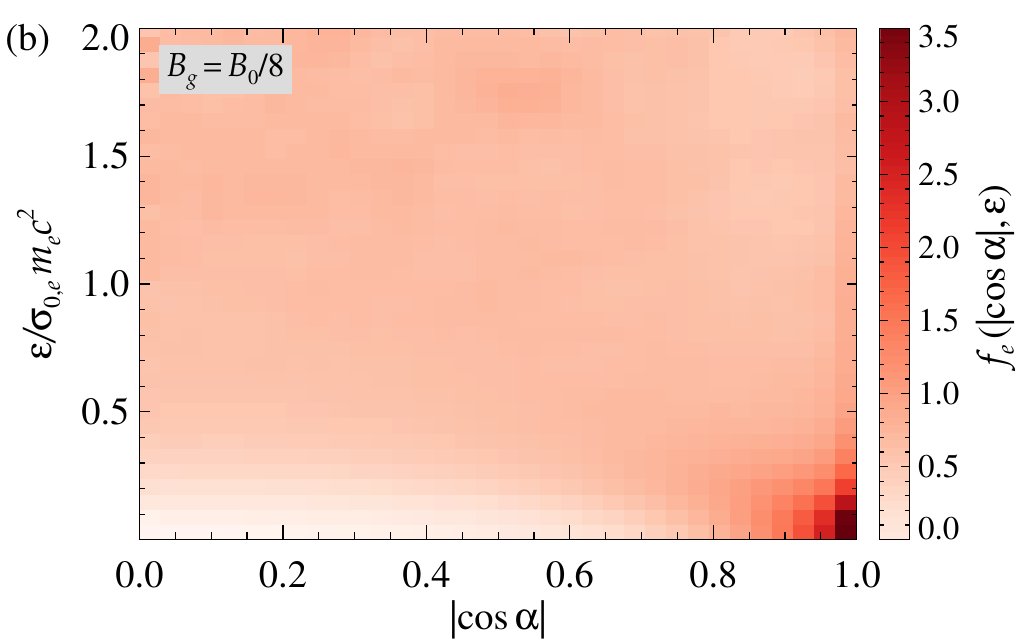} 
  \hfill
  \includegraphics[width=8.65cm]{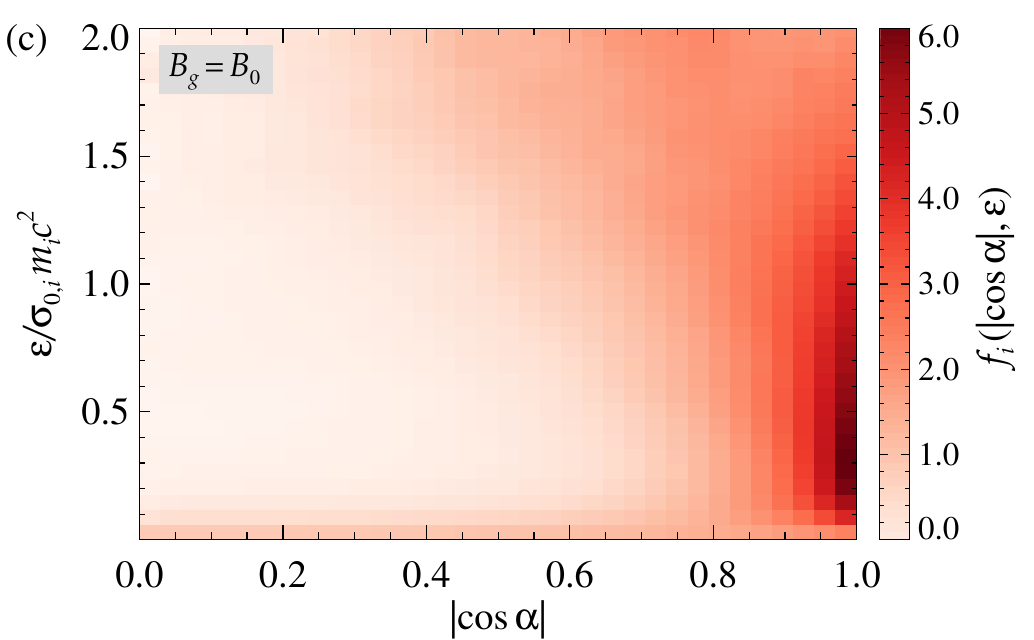} \hspace{0.3cm} 
  \includegraphics[width=8.65cm]{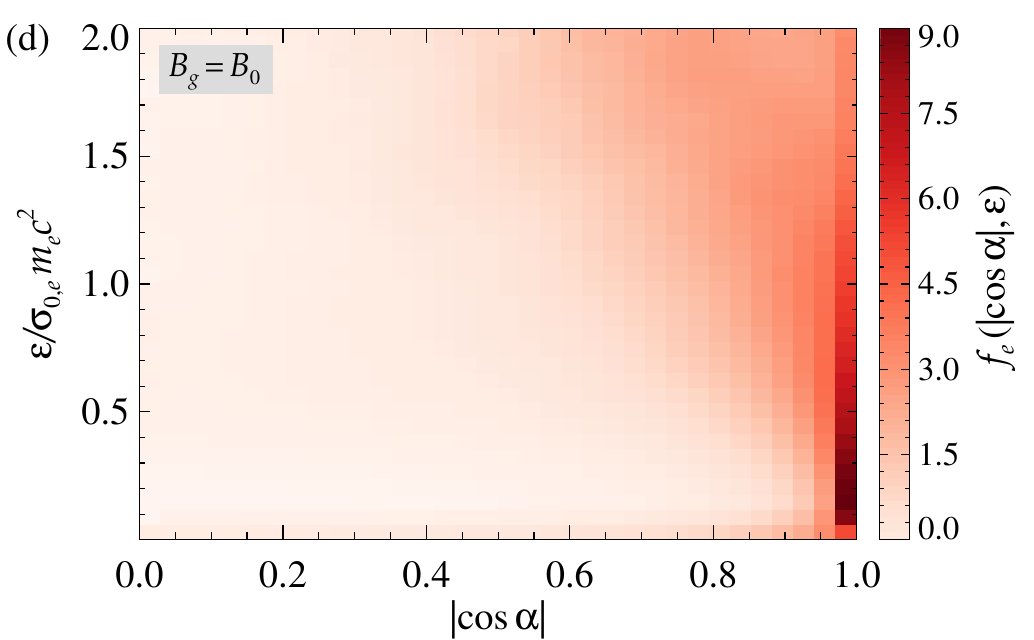} 
  \hfill
  \includegraphics[width=8.65cm]{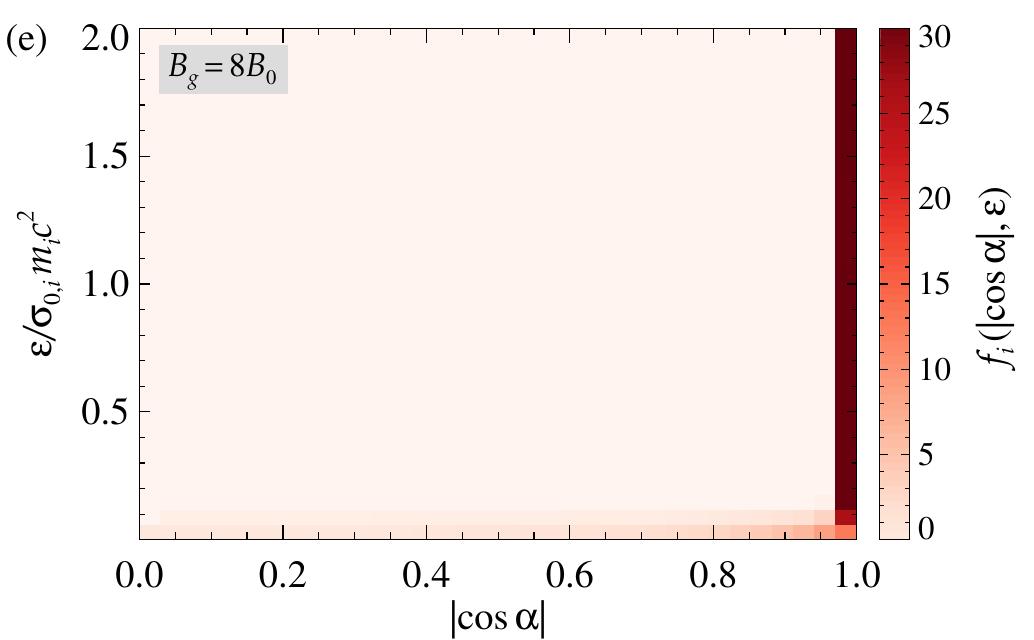} \hspace{0.3cm} 
  \includegraphics[width=8.65cm]{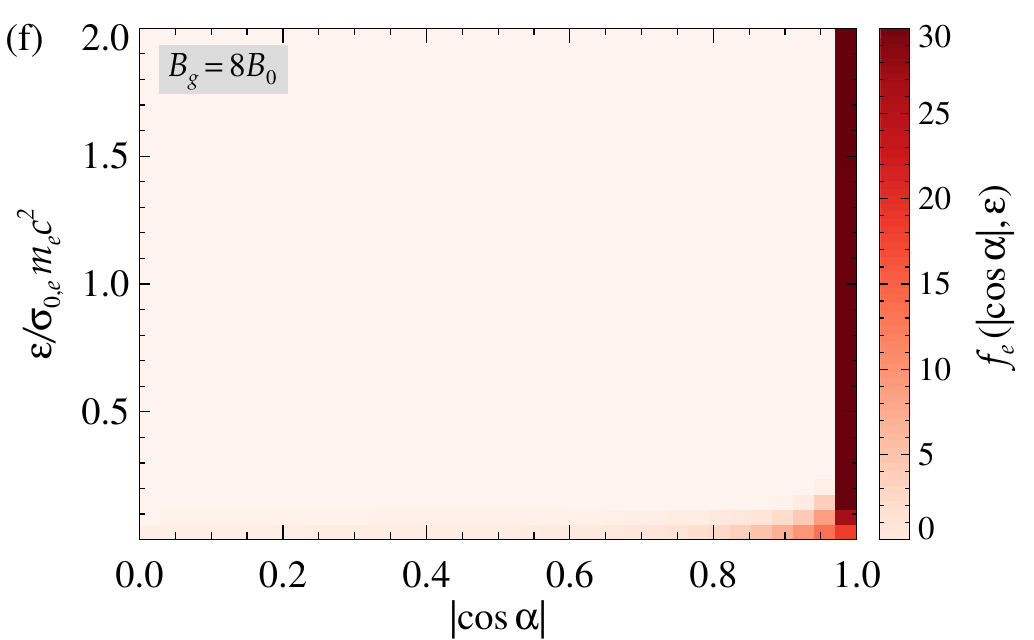} 
\caption{Particle distributions as a function of pitch-angle cosine $|\cos \alpha|$ and particle kinetic energy $\varepsilon$, at $t \sim 3 L_x/\langle v_{\rm out} \rangle$, for (a, b) $B_g=B_0/8$, (c, d) $B_g=B_0$, and (e, f) $B_g=8B_0$, alongside $\sigma_0=1$. The left column displays ion distributions, while the right column corresponds to electron distributions. $f_s \left( {|\cos \alpha|,\varepsilon} \right)$ is normalized such that ${\int_{0}^1  {f_s \left( {|\cos \alpha|,\varepsilon} \right) d(|\cos \alpha|)}}  = 1$. The $y$ axis is normalized by the characteristic energy scale $\sigma_{0,i} m_i c^2 = \sigma_{0,e} m_e c^2$ proportional to  the average particle energy of the energized plasma in the reconnection layer.} 
\label{2D_PA}
\end{figure*} 

Particle energization through magnetic reconnection results in anisotropic pitch angle distributions \citep{Comisso19,Comisso21,Comisso22,Comisso20ApJL,Comisso23}. 
In Figure \ref{2D_PA}, we present the distributions $f_s \left( {|\cos \alpha|,\varepsilon} \right)$ for ions ($s=i$) and electrons ($s=e$) from the same simulations depicted in Fig. \ref{fig2}. 
Here, $\alpha$ denotes the particle pitch angle, i.e. the angle between the particle momentum and the local magnetic field, while $\varepsilon = (\gamma - 1) m_s c^2$ denotes the particle kinetic energy, with $\gamma=(1-v^2/c^2)^{-1/2}$ indicating the particle Lorentz factor. 
We computed $\alpha$ and $\varepsilon$ in the local ${\bm{E}} \times {\bm{B}}$ frame, which is well-suited for evaluating radiative emission \citep{RybLig79}, as discussed in Section \ref{sec:discussion}. The distributions $f_s \left( {|\cos \alpha|,\varepsilon} \right)$ are normalized such that ${\int_{0}^1 {f_s \left( {|\cos \alpha|,\varepsilon} \right) d(|\cos \alpha|)}} = 1$, facilitating the identification of characteristic pitch angles across different particle energies. 
For all guide field strengths, both ions (left column) and electrons (right column) exhibit crests in $f_s \left( {|\cos \alpha|,\varepsilon} \right)$ at $|\cos \alpha| = 1$, indicating particles moving along the magnetic field lines. These peaks are a signature of particle ``injection'' \citep{Comisso19,Comisso21}—the rapid energization phase that extracts particles from the thermal pool and accelerates them to much higher energies. 

The characteristic features of the distributions $f_s \left( {|\cos \alpha|,\varepsilon} \right)$ vary depending on the ratio $B_g/B_0$. For $B_g/B_0 \ll 1$ (Figs. \ref{2D_PA}(a) and \ref{2D_PA}(b)), the crests at $|\cos \alpha| = 1$ reside at $\varepsilon \lesssim \sigma_{0,s} m_s c^2$, where $\sigma_{0,s} = {B_0^2}/{4\pi n_{s0} m_s c^2}$, while pitch angles tend towards isotropy with increasing particle energies. 
A similar pattern emerges for $B_g/B_0 \sim 1$ (Figs. \ref{2D_PA}(c) and \ref{2D_PA}(d)), albeit with more pronounced crests at $|\cos \alpha| = 1$ and a more gradual transition toward isotropy. 
On the other hand, for $B_g/B_0 \gg 1$ (Figs. \ref{2D_PA}(e) and \ref{2D_PA}(f)), the crests at $|\cos \alpha| = 1$ extend to the highest particle energies without displaying any tendency towards isotropy.

The emergence of pitch angle anisotropy can be understood in terms of a two-stage process, in line with previous studies of magnetic reconnection within a turbulence cascade \citep{Comisso18,Comisso19,Comisso21,Comisso22}. 
Initially, particles undergo acceleration by the reconnection electric field along the magnetic field on a timescale $t_{\rm init} \sim  \Delta\gamma {m_s c}/e \langle E_{\rm rec} \rangle \sim (\Delta\gamma/\epsilon_{\rm rec}) \langle c/v_{\rm out} \rangle \omega_{c,s}^{-1}$, where we assumed $v_{\parallel} \sim c$ and defined $\omega_{c,s}={e B_0}/{m_s c}$. This acceleration phase is potentially followed by pitch-angle scattering induced by magnetic field fluctuations, which acts to isotropize the distribution of particle pitch angles on the timescale $t_{\rm scatt} \sim (1 + B_g^2/B_0^2)L_x/v_{A0}$. Particle scattering, while tending to isotropize the pitch-angle distribution, also induces stochastic particle acceleration ($t_{\rm acc} \propto t_{\rm scatt})$. However, as $B_g/B_0$ increases, $t_{\rm scatt}$ becomes significantly larger than $L_x/\langle v_{\rm out} \rangle$, resulting in the progressive suppression of both pitch angle isotropization and stochastic particle acceleration.

\begin{figure*}
 \centering 
  \includegraphics[width=8.65cm]{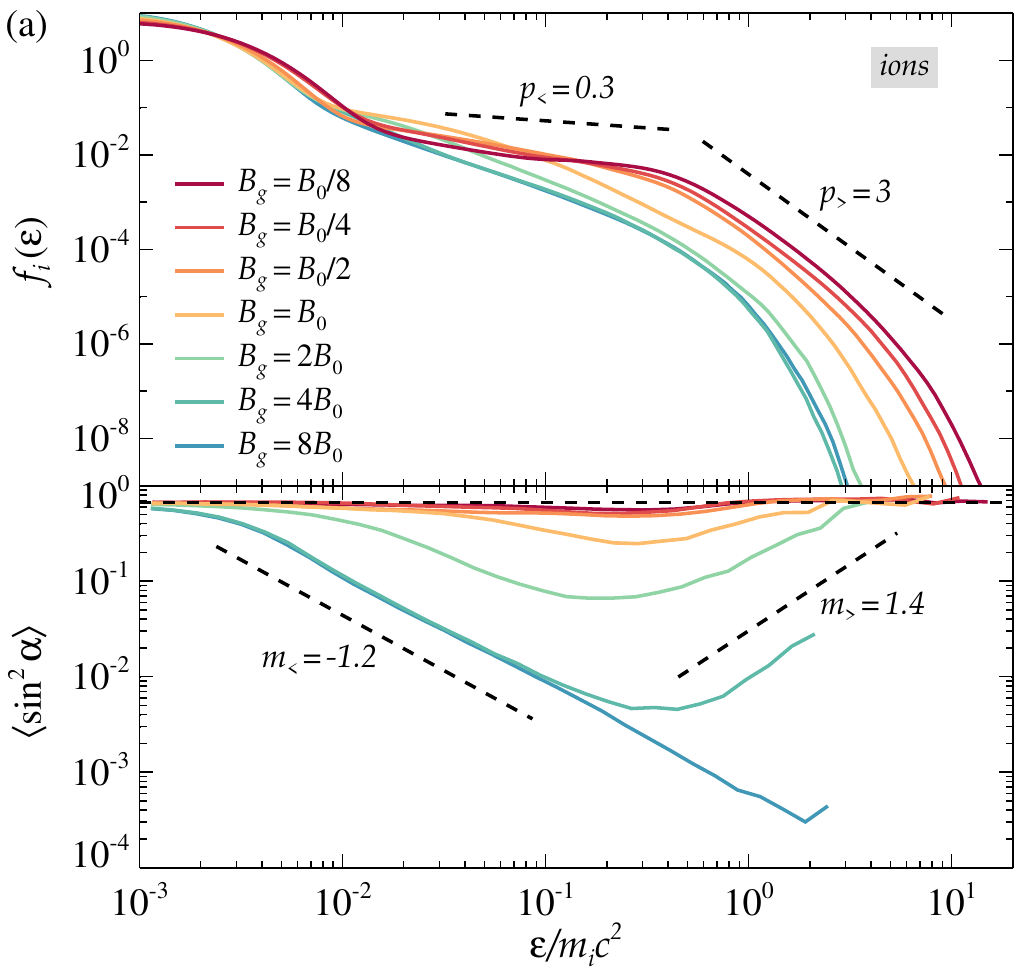}
  \includegraphics[width=8.65cm]{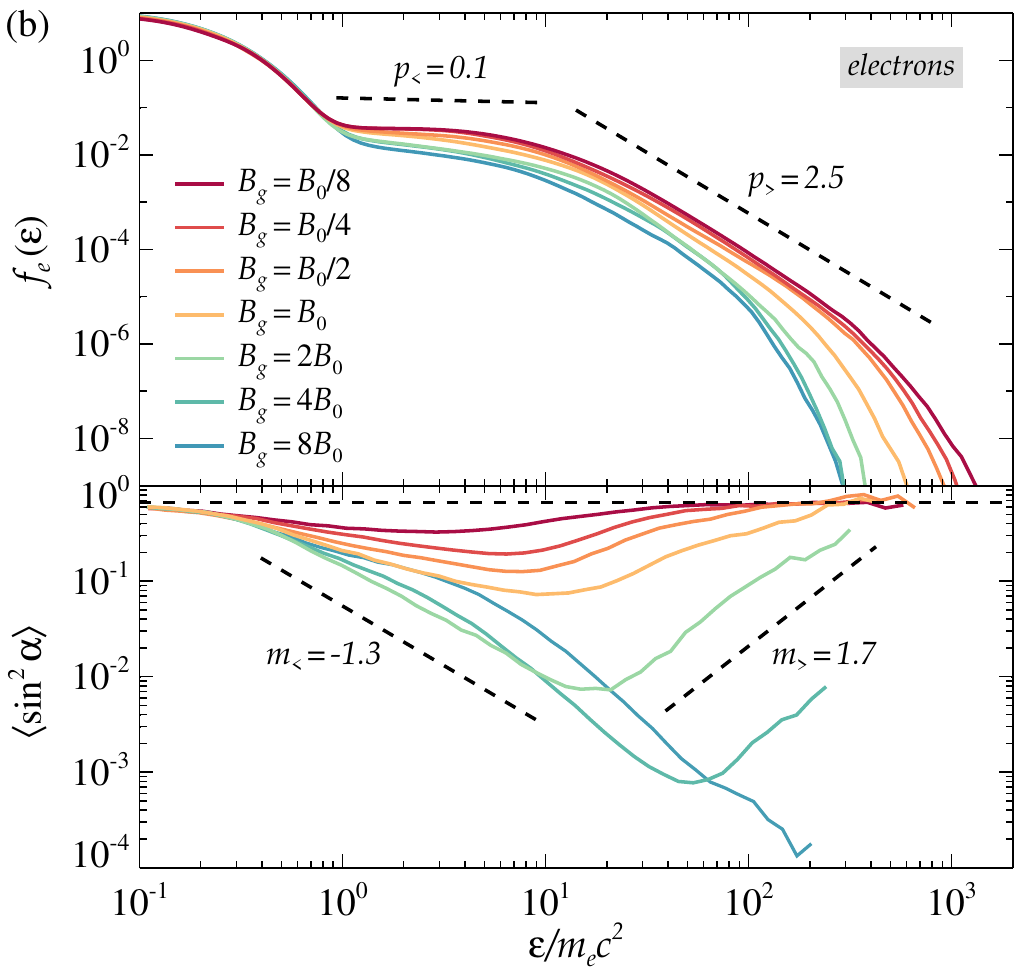}
\caption{(a) Ion and (b) electron energy spectra, $f_s (\varepsilon) = dN(\varepsilon)/d\varepsilon$, and mean value of $\sin^2 \alpha$ as a function of $\varepsilon$, obtained from simulations with different guide field strengths $B_g = (1/8, 1/4, 1/2, 1, 2, 4, 8) B_0$, with fixed magnetization $\sigma_0 = 1$. 
The particle energy spectra and pitch angle anisotropy are integrated over the entire reconnection region ($|y| \leq 0.25 L_x$) and evaluated at late times, $t \simeq 3 L_x/\langle v_{\rm out} \rangle$, when they have fully developed. Dashed lines indicating power-law slopes are provided for reference. In the bottom frames, a horizontal dashed black line indicates the expected value for an isotropic pitch angle distribution, $\langle \sin^2 \alpha \rangle = 2/3$.}
\label{fig4}
\end{figure*} 

\begin{figure*}
 \centering 
  \includegraphics[width=8.65cm]{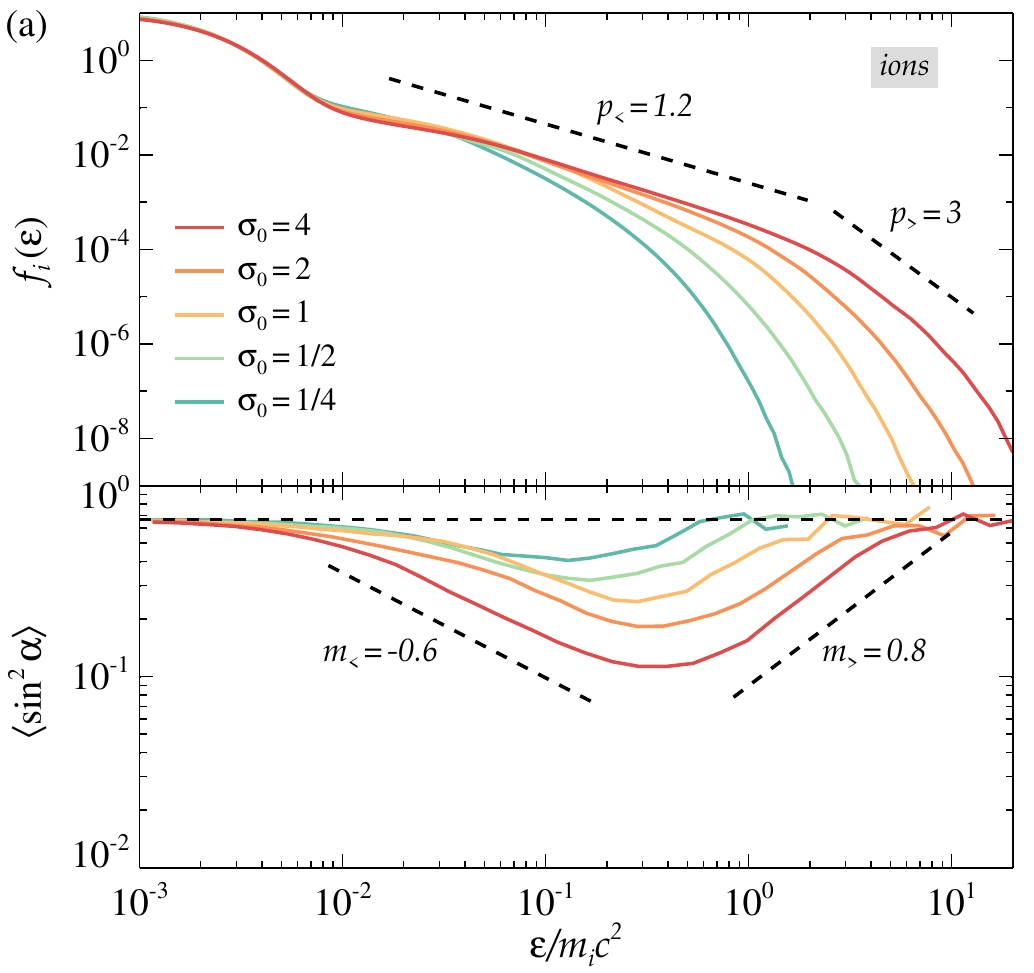}
  \includegraphics[width=8.65cm]{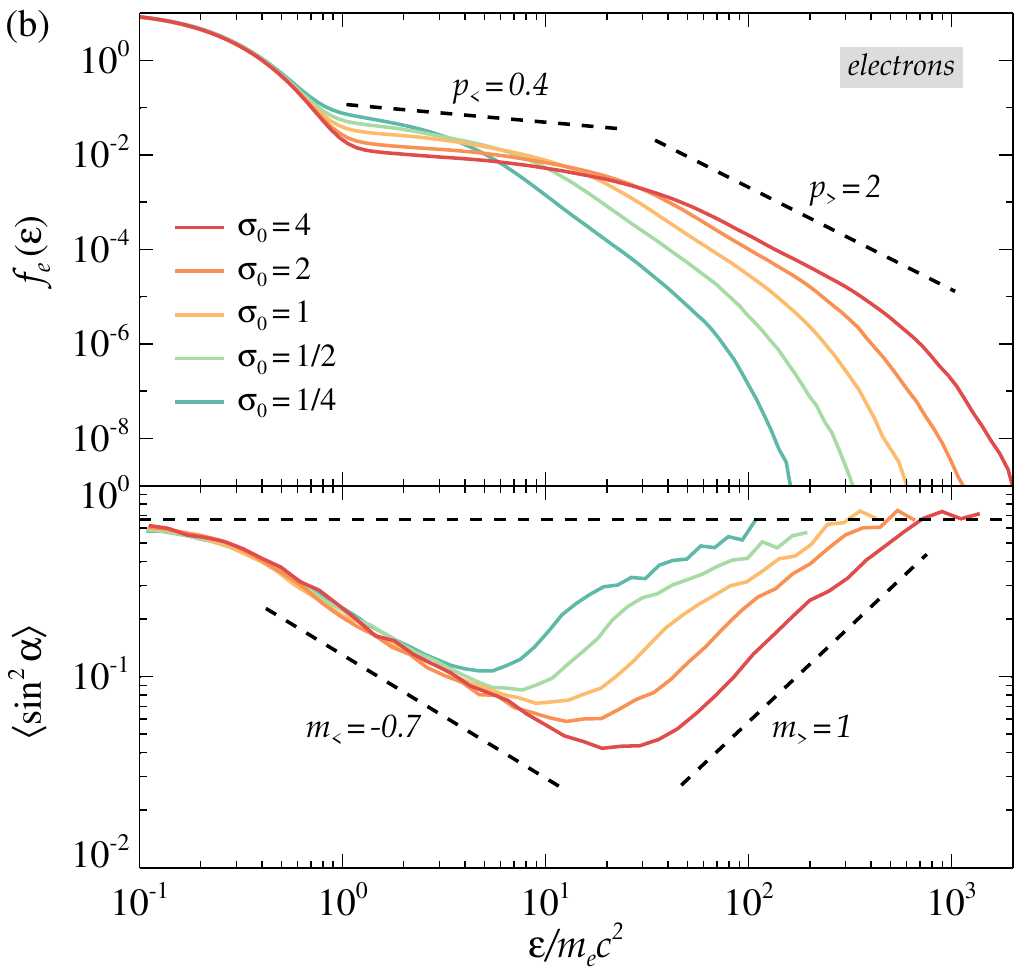}
\caption{(a) Ion and (b) electron energy spectra, $f_s (\varepsilon) = dN(\varepsilon)/d\varepsilon$, and mean value of $\sin^2 \alpha$ as a function of $\varepsilon$, obtained from simulations with different magnetization $\sigma_0 = (1/4, 1/2, 1, 2, 4)$, with fixed guide field strength $B_g/B_0 = 1$. 
The particle energy spectra and pitch angle anisotropy are integrated over the entire reconnection region ($|y| \leq 0.25 L_x$) and evaluated at late times, $t \simeq 3 L_x/\langle v_{\rm out} \rangle$, when they have fully developed.
Dashed lines indicating power-law slopes are provided for reference. In the bottom frames, a horizontal dashed black line indicates the expected value for an isotropic pitch angle distribution, $\langle \sin^2 \alpha \rangle = 2/3$.}
\label{fig5}
\end{figure*} 

The influence of the guide field on particle acceleration and the generation of anisotropic pitch-angle distributions is illustrated in Figure \ref{fig4}, focusing on cases with $\sigma_0 = 1$. The upper frames present ion and electron energy spectra, $f_s (\varepsilon) = dN(\varepsilon)/d\varepsilon$, while the lower frames display the corresponding mean squared pitch angle sine for particles at given kinetic energy, $\langle \sin^2 \alpha \rangle$.  
Analogous measures for simulations with varying magnetization and $B_g/B_0=1$ are provided in Figure \ref{fig5}. Particle energy spectra and pitch angle anisotropy are evaluated at a developed stage, approximately $t_{\rm dev} \simeq 3 L_x/\langle v_{\rm out} \rangle$ from the onset of fast reconnection. A distinct characteristic of both $f_s (\varepsilon) = dN(\varepsilon)/d\varepsilon$ and $\langle \sin^2 \alpha \rangle$ is their broken power-law structure. 

\begin{figure}
\begin{center}
    \includegraphics[width=8.65cm]{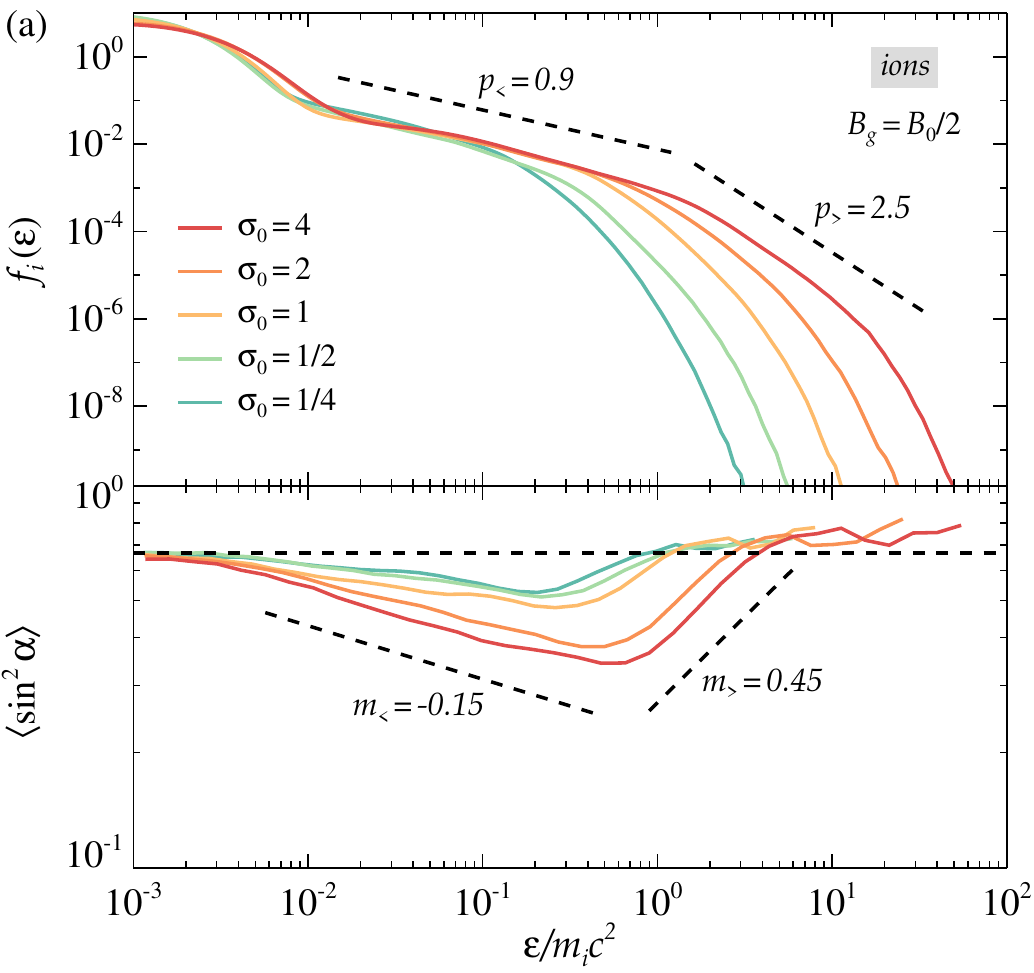}
\vspace{0.3cm}
    
    \includegraphics[width=8.65cm]{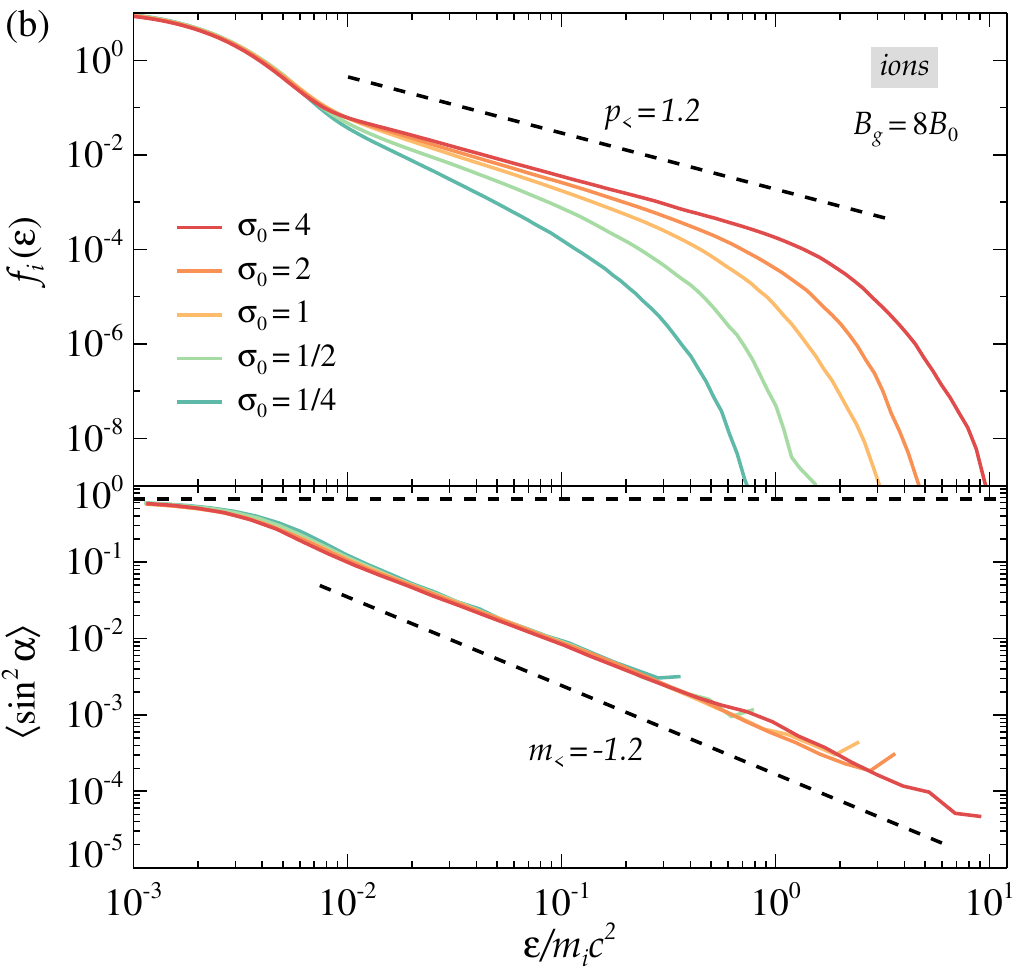}
\end{center}
\vspace{-0.2cm}
    \caption{Ion energy spectra, $f_i (\varepsilon) = dN(\varepsilon)/d\varepsilon$, and mean value of $\sin^2 \alpha$ as a function of $\varepsilon$, from simulations with magnetizations $\sigma_0 = (1/4, 1/2, 1, 2, 4)$ and guide field strengths (a) $B_g = B_0/2$ and (b) $B_g = 8 B_0$. 
The particle energy spectra and pitch angle anisotropy are integrated over the entire reconnection region ($|y| \leq 0.25 L_x$) and evaluated at late times, $t \simeq 3 L_x/\langle v_{\rm out} \rangle$, as in Figs. \ref{fig4} and \ref{fig5}.
Dashed lines indicating power-law slopes are provided for reference. In the bottom frames, a horizontal dashed black line indicates the expected value for an isotropic pitch angle distribution, $\langle \sin^2 \alpha \rangle = 2/3$.}
\label{fig6}
\end{figure}

In both Figs. \ref{fig4} and \ref{fig5}, the spectra $f_s (\varepsilon) = dN(\varepsilon)/d\varepsilon$ for both ions (panels (a)) and electrons (panels (b)) exhibit extremely hard power-law slopes ($p_< \lesssim 1$) in the lower-energy (injection) range, while steepening ($p_> \gtrsim 2$) beyond this range. The slopes below injection are consistently hard, whereas their values above injection depend significantly on $B_g/B_0$ and $\sigma_0$ \citep[e.g.][]{XLi2017,Werner2017ApJL,Ball18,Hakobyan21,LiX23,Comisso23,Zhang23ApJL,French23}.  Higher $B_g/B_0$ and lower $\sigma_0$ result in a steeper spectrum above injection, potentially leading to the disappearance of the high-energy range. For the values of $B_g/B_0$ and $\sigma_0$ considered in Figs. \ref{fig4} and \ref{fig5}, this occurs for the ions when $B_g/B_0 \gtrsim 1$ and $\sigma_0 \lesssim 1$. Further evidence is provided in Fig. \ref{fig6}, illustrating the ion energy spectra for varying $\sigma_0$ and guide field strengths of $B_g = B_0/2$ (Fig. \ref{fig6}(a)) and $B_g = 8 B_0$ (Fig. \ref{fig6}(b)). The higher-energy range with $p_> \simeq 2.5$ for $\sigma_0 = 4$ and $B_g = B_0/2$ vanishes when the guide field increases to $B_g = 8 B_0$.

\begin{figure}
\begin{center}
    \includegraphics[width=8.65cm]{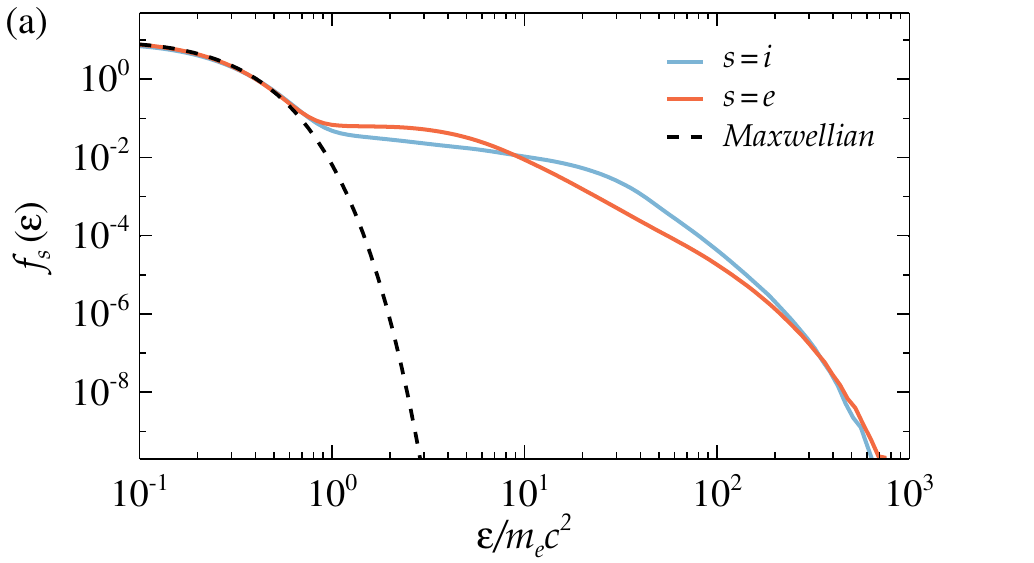}
\vspace{0.3cm}
    \includegraphics[width=8.65cm]{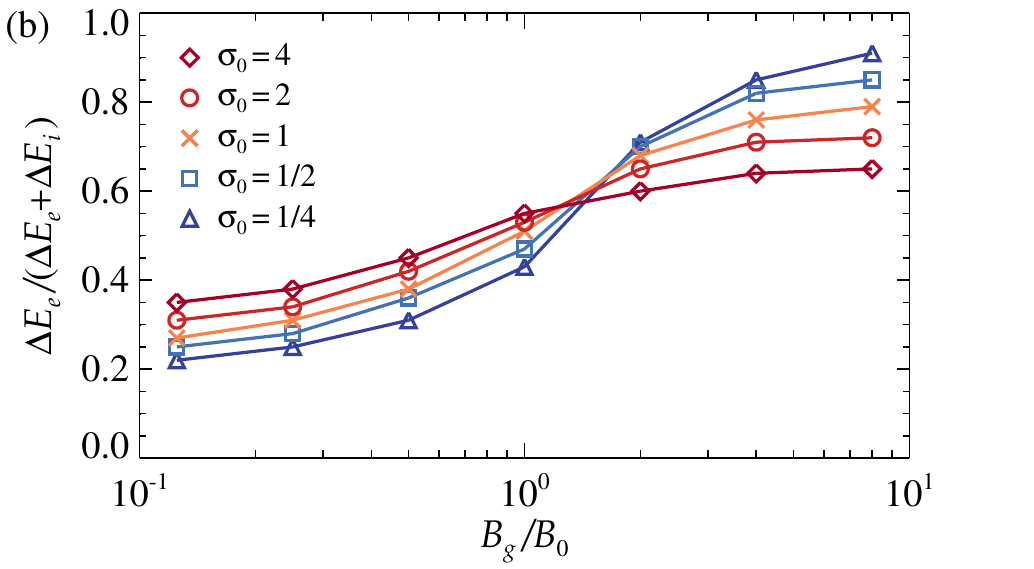}
\vspace{0.3cm}        
    \includegraphics[width=8.65cm]{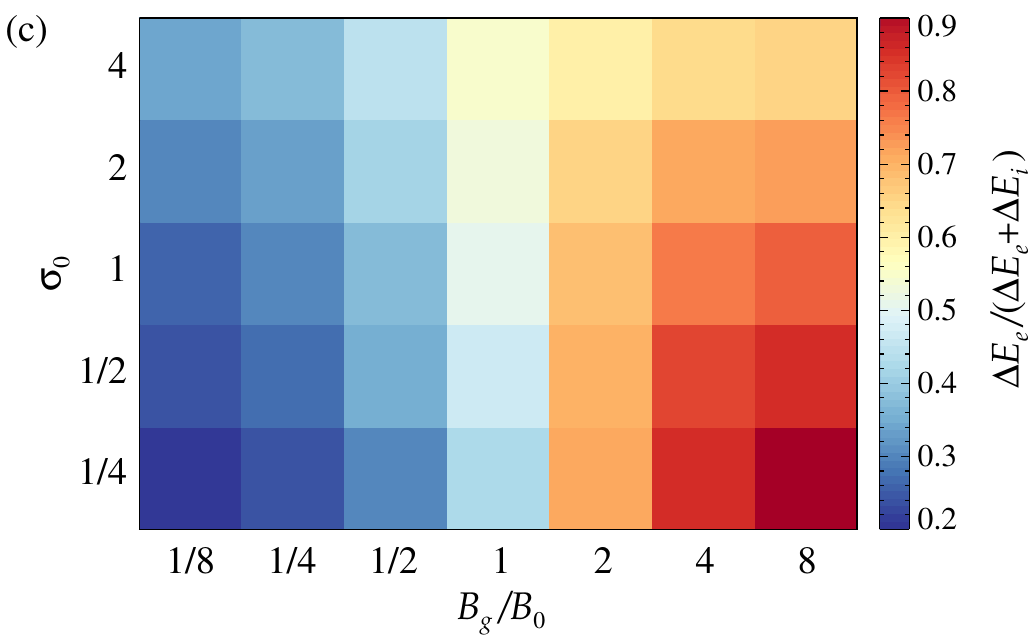}
\end{center}
\vspace{-0.2cm}
    \caption{(a) Ion (blue line) and electron (red line) energy spectra at a late time, along with a Maxwellian fit (black dashed line), for a simulation with $B_g=B_0/4$ and $\sigma_0=1/2$.
(b) Partitioning of the released magnetic energy into electrons vs. ions, indicated by the fraction $\Delta{E_e}/(\Delta{E_e}+\Delta{E_i})$, as a function of $B_g/B_0$ for different values of the magnetization $\sigma_0$. Symbol markers correspond to $\sigma_0 = 4$ ({\large$\diamond$} in dark red), $\sigma_0 = 2$ ({\large$\circ$} in light red), $\sigma_0 = 1$ ($\times$ in orange), $\sigma_0 = 1/2$ ($\square$ in light blue), and $\sigma_0 = 1/4$ ($\vartriangle$ in dark blue). 
(c) 2D histogram illustrating the variation of the ratio $\Delta{E_e}/(\Delta{E_e}+\Delta{E_i})$ across the parameter space of $B_g/B_0$ and $\sigma_0$.}
\label{fig7}
\end{figure}

The lower panels of Figs. \ref{fig4} and \ref{fig5} depict the energy dependence of $\langle \sin^2 \alpha \rangle$ for both ions and electrons. 
Electrons exhibit higher levels of anisotropy, as evident from the deviation of $\min \langle \sin^2 \alpha \rangle$ from the isotropic expectation ($\langle \sin^2 \alpha \rangle = 2/3$, denoted by the horizontal dashed black lines). 
As the magnetization increases, deviations from isotropy extend to higher Lorentz factors. Specifically, the range of Lorentz factors affected by the pitch-angle anisotropy scales linearly with the magnetization associated with the relevant plasma species, $\sigma_{0,s}$ (note that since we employed a reduced mass ratio $m_i/m_e=100$ in our PIC simulations, this results in artificially lower Lorentz factors for electrons in Figs. \ref{fig4}(b) and \ref{fig5}(b) compared to a realistic proton-to-electron mass ratio $m_i/m_e \simeq 1836$).
As the guide field strengthens, the power-law slopes $m_<$ and $m_>$ characterizing $\langle \sin^2 \alpha \rangle$ become steeper. Eventually, when the guide field falls well within the regime $B_g/B_0 \gg 1$, the power-law range with a positive slope $m_>$ disappears, as in this regime, high-energy particles cannot undergo effective pitch angle isotropization and further acceleration \citep{Comisso23}. This behavior is observed for the case $B_g = 8 B_0$ in Figure \ref{fig4}.
Further confirmation is provided in Fig. \ref{fig6}, illustrating $\langle \sin^2 \alpha \rangle$ for ions under guide field strengths $B_g = B_0/2$ (Fig. \ref{fig6}(a)) and $B_g = 8 B_0$ (Fig. \ref{fig6}(b)). 
Only modest deviations from anisotropy are present for $B_g = B_0/2$, whereas for $B_g = 8 B_0$, as the anisotropy becomes notably strong, the positive slope $m_>$ disappears for any $\sigma_0$. 

The ratio $B_g/B_0$ also governs which particle species predominantly receives the magnetic energy released during reconnection. 
We assess this by evaluating the excess energy beyond the Maxwellian of the ambient plasma for both ions and electrons. 
In Figure \ref{fig7}(a), we present an example of ion (blue line) and electron (red line) energy spectra, alongside the fitting of the Maxwellian distribution of the ambient plasma.  
The reconnection process results in electrons gaining energy $\Delta{E_e}$ and ions acquiring energy $\Delta{E_i}$. 
In Figure \ref{fig7}(b), we show the fraction of energy gained by electrons relative to the total population of particles, $\Delta{E_e}/(\Delta{E_e}+\Delta{E_i})$, as a function of $B_g/B_0$, across various $\sigma_0$ values. 
For $B_g/B_0 \ll 1$, ions gain more energy than electrons \citep[see also][]{Rowan17,Werner18}. 
However, as the guide field strength increases, energy is redirected towards electrons, resulting in $\Delta{E_e} \sim \Delta{E_i}$ for $B_g/B_0 \sim 1$ and higher energy for electrons rather than ions when $B_g/B_0 \gg 1$ \citep[see also][]{Rowan19,Werner24}.
The energy gain difference between electrons and ions diminishes as $\sigma_0$ increases. For $\sigma_0$ rising well above unity, both species attain ultrarelativistic energies, approaching a pair-plasma behavior. Consequently, for $\sigma_0 \gg 1$, $\Delta{E_e} \sim \Delta{E_i}$ irrespective of $B_g/B_0$. A 2D histogram highlighting the discussed trend is provided in Figure \ref{fig7}(c). 

\begin{figure}
    \includegraphics[width=8.15cm]{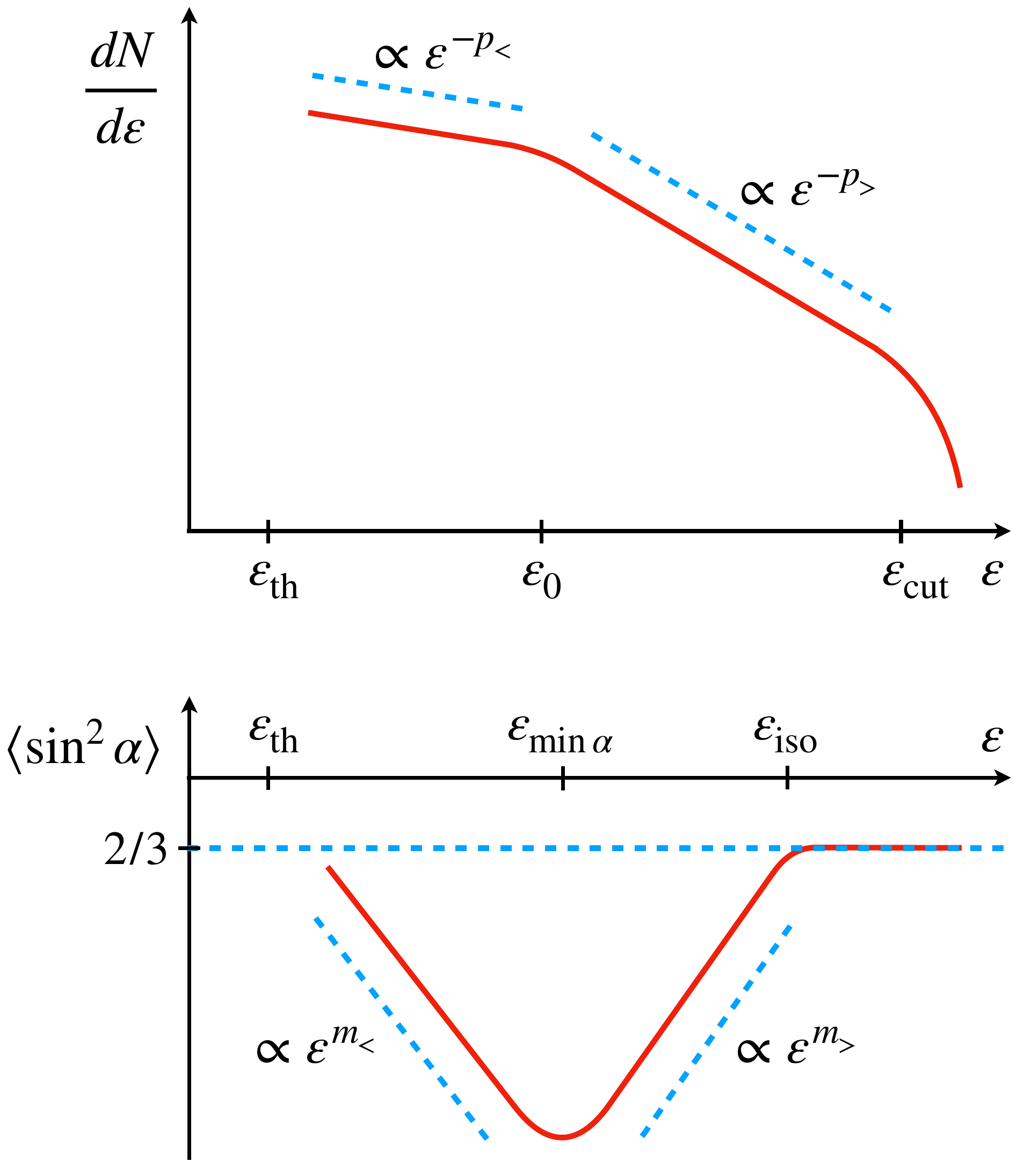}
\vspace{-0.2cm}
    \caption{Illustration of the characteristic broken power laws for $f (\varepsilon) = dN(\varepsilon)/d\varepsilon$ (top frame) and $\langle \sin^2 \alpha \rangle$ (bottom frame) resulting from magnetic reconnection. 
$\varepsilon_0$ and $\varepsilon_{\min \alpha}$ indicate the break kinetic energies for the particle energy spectrum and the mean squared pitch angle sine, respectively. 
$p_<$ and $p_>$ denote the (negative) slopes of the particle energy spectrum below and above $\varepsilon_0$, respectively.
$m_<$ and $m_>$ denote negative and positive slopes below and above $\varepsilon_{\min \alpha}$, respectively.
$\varepsilon_{\rm iso}$ indicates the kinetic energy at which particles return to a state close to pitch-angle isotropy.
Finally, $\varepsilon_{\rm th}$ is the characteristic thermal energy, while $\varepsilon_{\rm cut}$ denotes the cutoff energy for the higher-energy power-law range.}
\label{figsketch}
\end{figure}

\begin{figure*}
 \centering 
  \includegraphics[width=8.65cm]{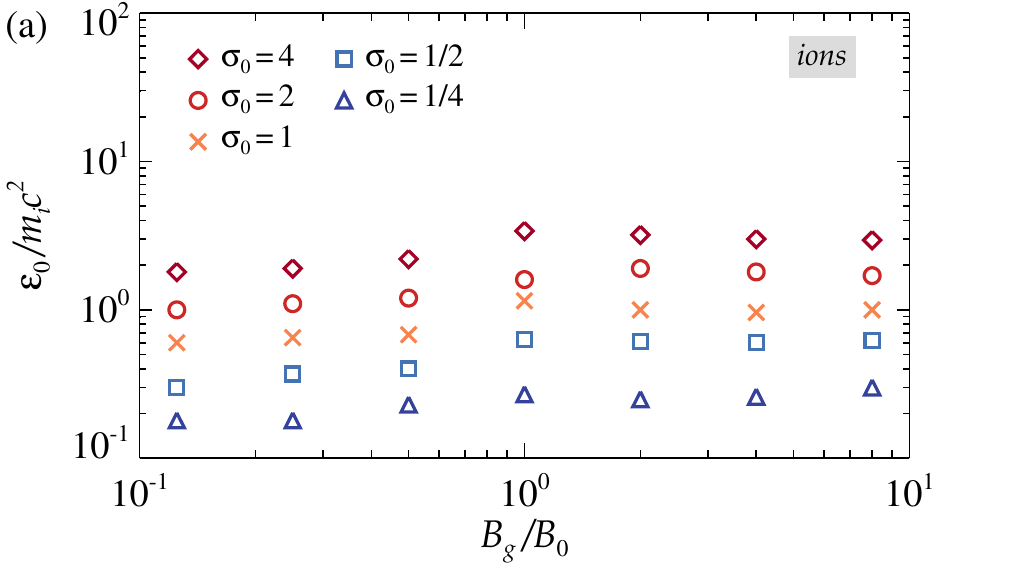}
  \includegraphics[width=8.65cm]{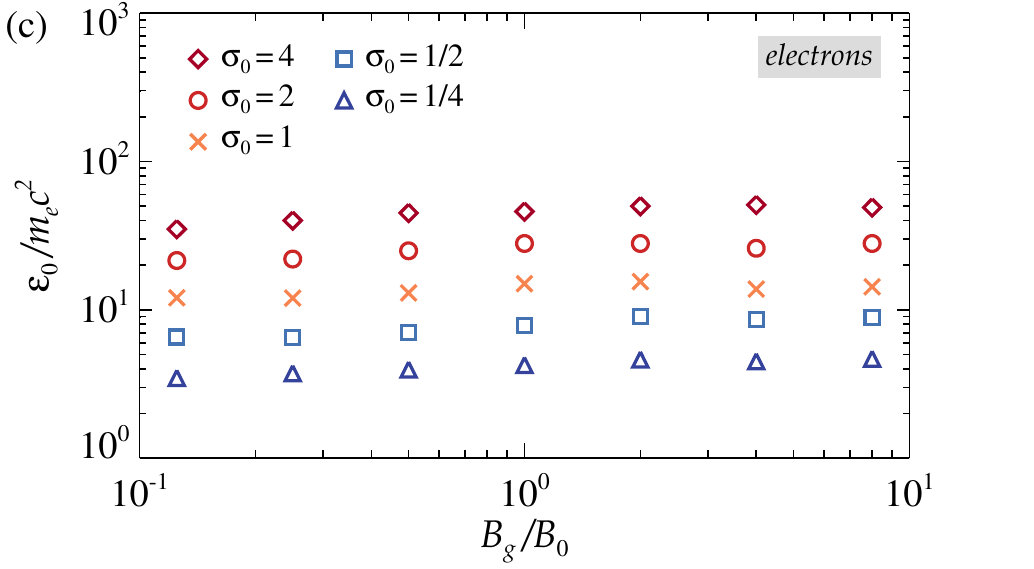}
  \hfill
  \includegraphics[width=8.65cm]{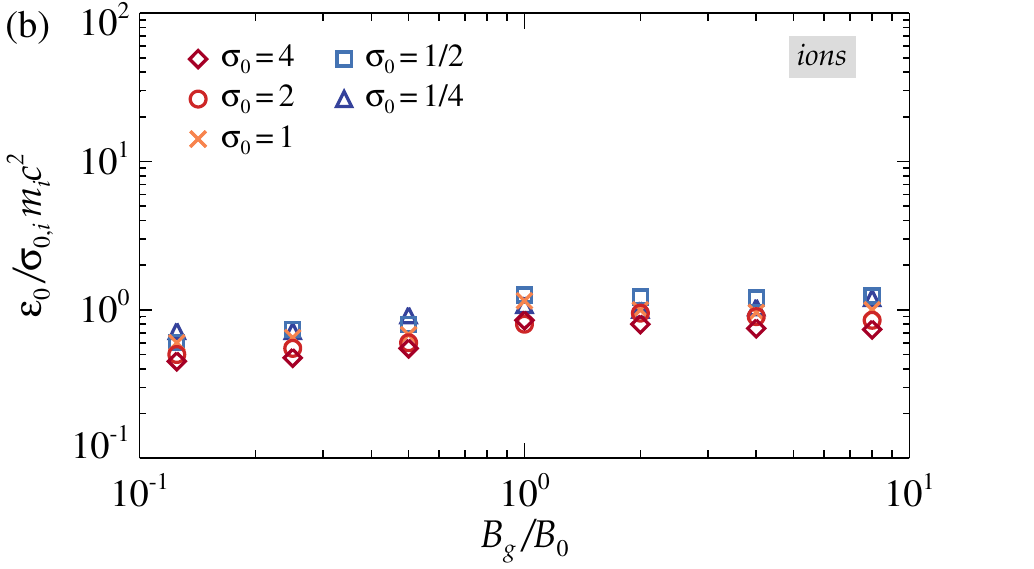}
  \includegraphics[width=8.65cm]{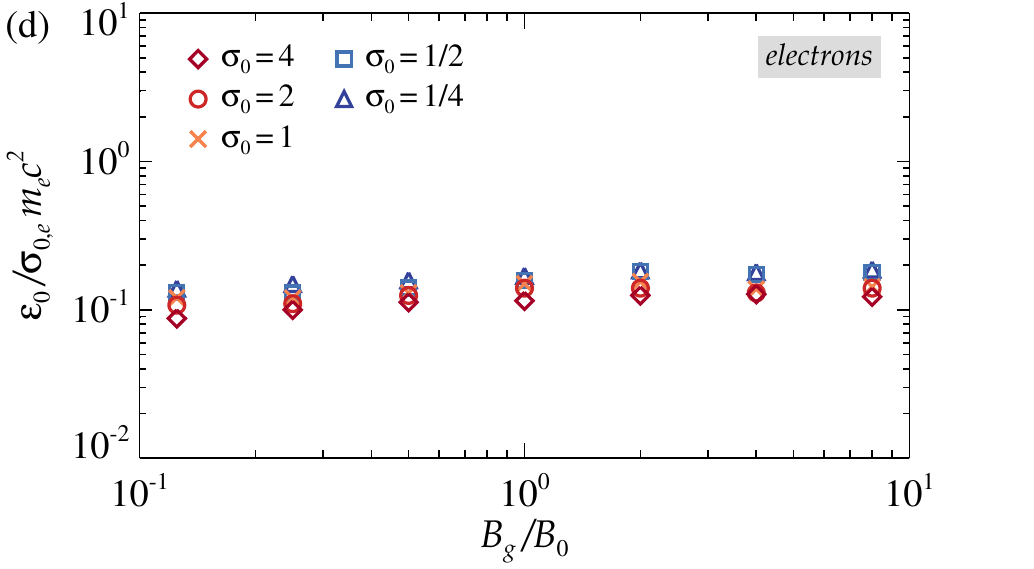}
\caption{Break particle kinetic energy $\varepsilon_0$ for ions (left column) and electrons (right column) for the different dimensionless guide field strengths $B_g/B_0$ and various plasma magnetizations $\sigma_0$. Symbols denote different values of $\sigma_0$: $\sigma_0 = 4$ ({\large$\diamond$} in dark red), $\sigma_0 = 2$ ({\large$\circ$} in light red), $\sigma_0 = 1$ ($\times$ in orange), $\sigma_0 = 1/2$ ($\square$ in light blue), and $\sigma_0 = 1/4$ ($\vartriangle$ in dark blue). In the top row, $\varepsilon_0$ is normalized by $m_s c^2$ ($s=i$ for ions, $s=e$ for electrons), while in the bottom row, it is normalized by $\sigma_{0,i} m_i c^2 = \sigma_{0,e} m_e c^2$ to highlight the typical energy scale associated with $\varepsilon_0$.}
\label{fig9}
\end{figure*} 

In Figure \ref{figsketch}, we summarize the characteristics of $f (\varepsilon) = dN(\varepsilon)/d\varepsilon$ and $\langle \sin^2 \alpha \rangle$ observed from our simulation campaign. 
Both ions and electrons exhibit nonthermal energy spectra that can be described by a broken power law: 
\begin{equation}
\label{eq:dNdgamma}
{N(\varepsilon)}{d\varepsilon} = 
\begin{cases}
K (\varepsilon/\varepsilon_0)^{-p_<} {d\varepsilon}  \;  , &  \quad \varepsilon_{\rm th}<\varepsilon<\varepsilon_{0} \\ 
K (\varepsilon/\varepsilon_0)^{-p_>} {d\varepsilon} \;  , & \quad \varepsilon_{0}<\varepsilon<\varepsilon_{\rm cut} 
\end{cases}
\end{equation} 
where $\varepsilon_0$ is the break energy that separates the two power-law ranges, with $p_<$ and $p_>$ denoting the power-law indices for the lower and upper ranges, respectively. Additionally, $\varepsilon_{\rm th} = (3/2)k_B T_{0}$ and $\varepsilon_{\rm cut}$ denote the thermal energy and cutoff energy, while $K$ is a normalization constant. 

A broken power law is also observed in the mean squared pitch angle sine over specific energy ranges. This broken power law for $\langle \sin^2 \alpha \rangle$ can be described as follows: 
\begin{equation}
\label{eq:alphagamma}
\langle \sin^2 \alpha \rangle = 
\begin{cases}
\Lambda \left({\varepsilon}/{\varepsilon_{\min \alpha}}\right)^{m_<}  \;  , &  \quad \varepsilon_{\rm th}<\varepsilon<\varepsilon_{\min \alpha} \\ 
\Lambda \left({\varepsilon}/{\varepsilon_{\min \alpha}}\right)^{m_>}  \;  , &  \quad \varepsilon_{\min \alpha}<\varepsilon<\varepsilon_{\rm iso} \\ 
2/3 \;  , & \quad \varepsilon_{\rm iso}<\varepsilon<\varepsilon_{\rm cut} \, .
\end{cases}
\end{equation}
Here, $\varepsilon_{\min \alpha}$ represents the energy at which pitch-angle anisotropy is most pronounced,  $m_<$ and $m_>$ are the power-law indices characterizing the negative and positive energy dependencies of the pitch-angle anisotropy, respectively, $\Lambda \sim \min \langle \sin^2 \alpha \rangle$, and $\varepsilon_{\rm iso}$ denotes the energy at which particles return to a state close to pitch-angle isotropy.

\begin{figure*}
 \centering 
  \includegraphics[width=8.65cm]{Fig_ions}
  \includegraphics[width=8.65cm]{Fig_electrons}
  \hfill
  \includegraphics[width=8.65cm]{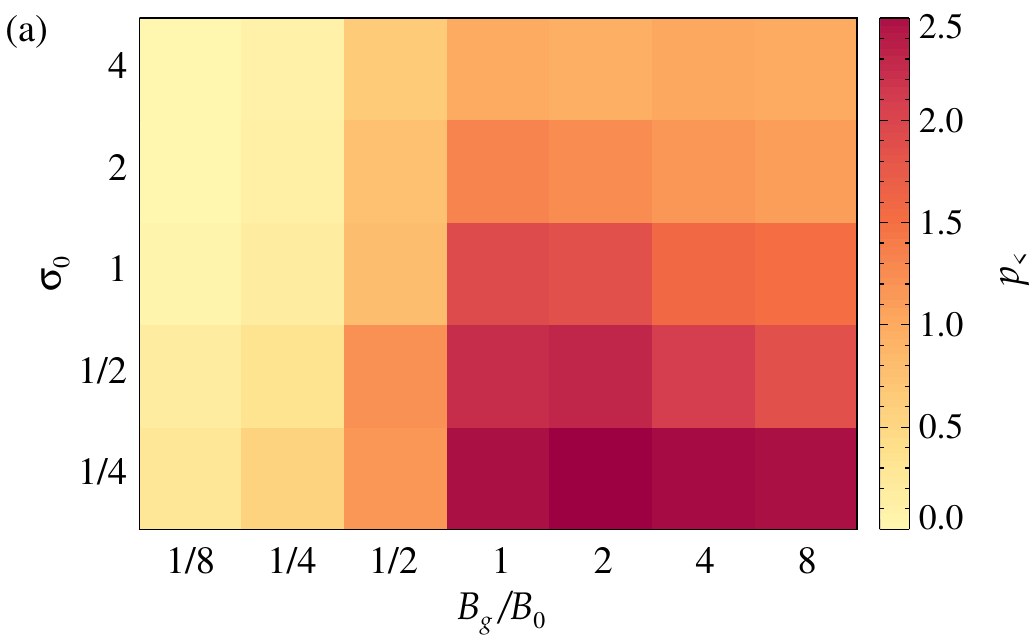}
  \includegraphics[width=8.65cm]{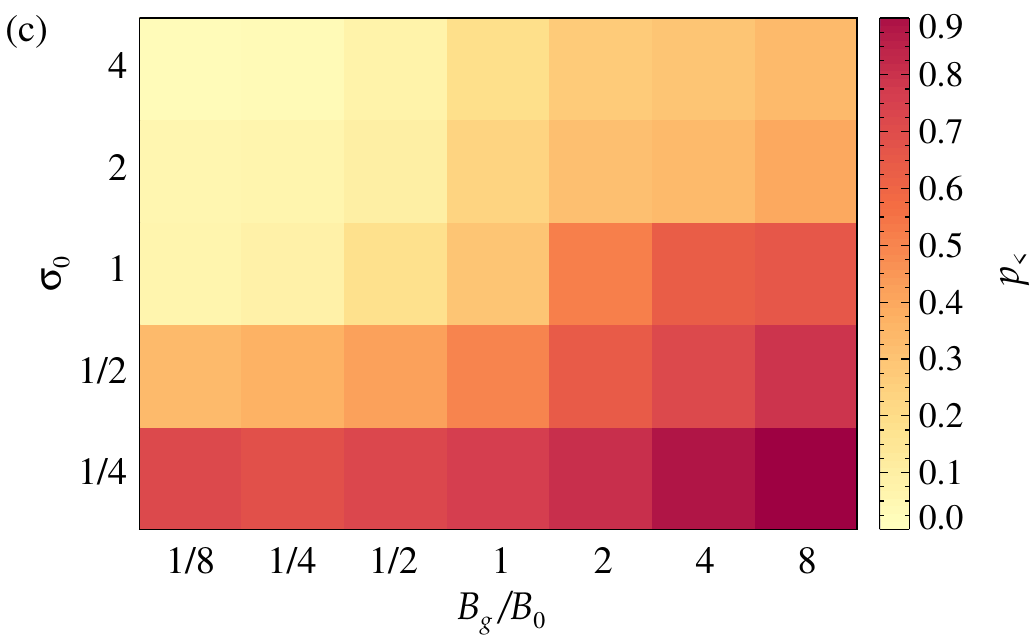}
  \hfill
  \includegraphics[width=8.65cm]{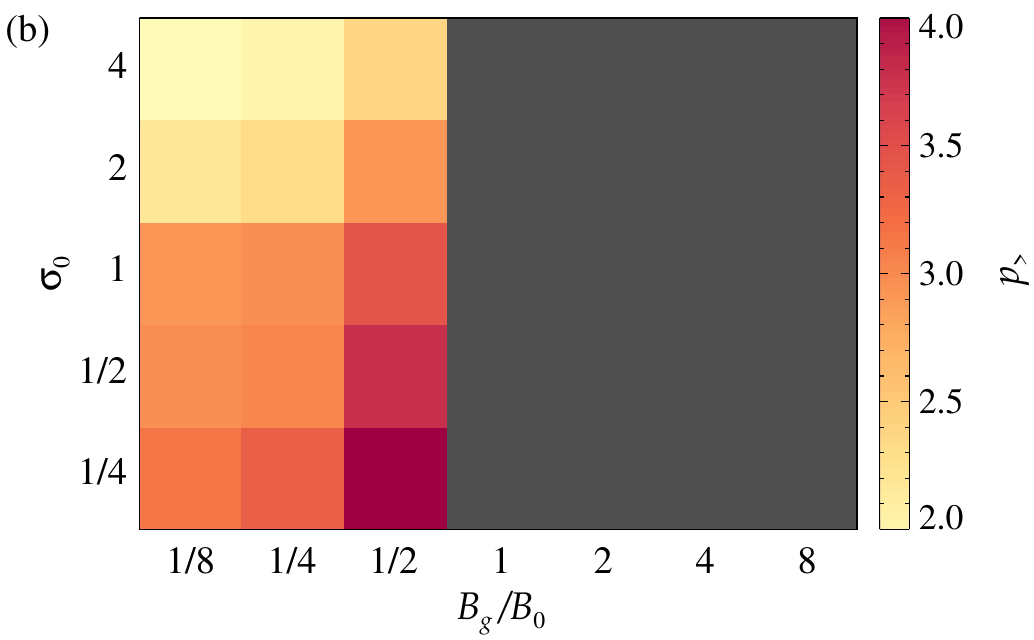}
  \includegraphics[width=8.65cm]{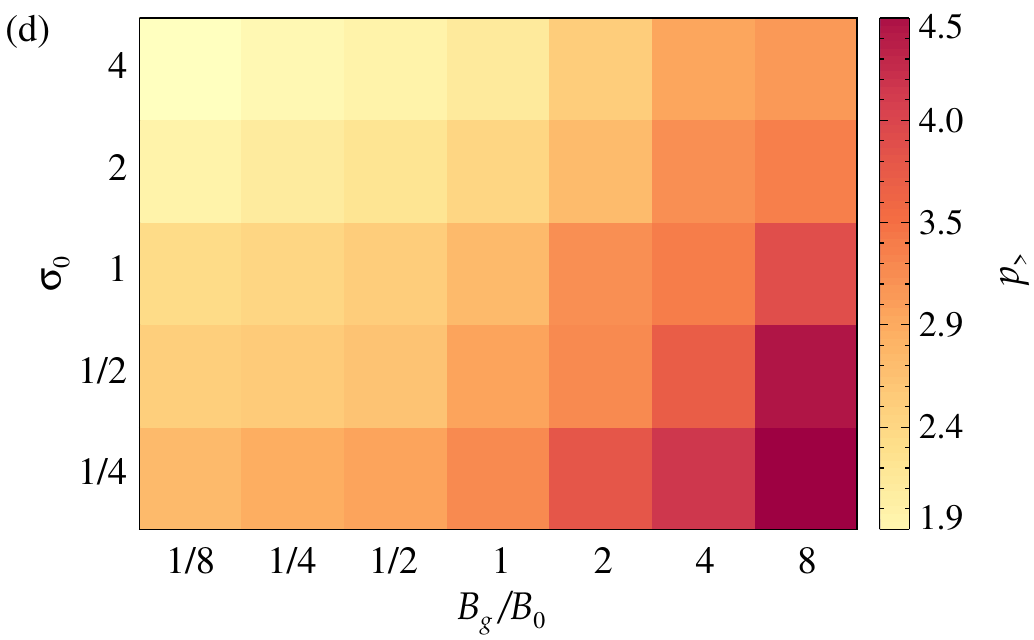}
\caption{2D histograms providing characterization of particle energy spectra for ions (left) and electrons (right) across different values of the dimensionless guide field strength $B_g/B_0$ and plasma magnetization $\sigma_0$. The top row (panels (a) and (c)) shows the power-law indices $p_<$, while the bottom row (panels (b) and (d)) presents the power-law indices $p_>$. The dark gray shading in panel (c) indicates that a power-law range with slope $p_>$ is not readily discernible or absent for the specified values of $B_g/B_0$ and $\sigma_0$.}
\label{fig10}
\end{figure*} 

\begin{figure*}
 \centering 
  \includegraphics[width=8.65cm]{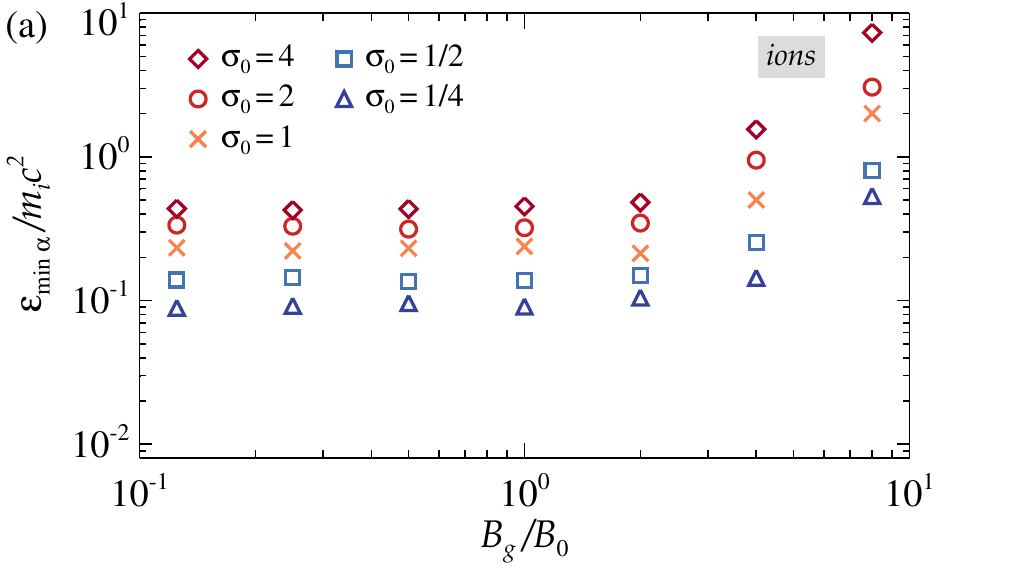}
  \includegraphics[width=8.65cm]{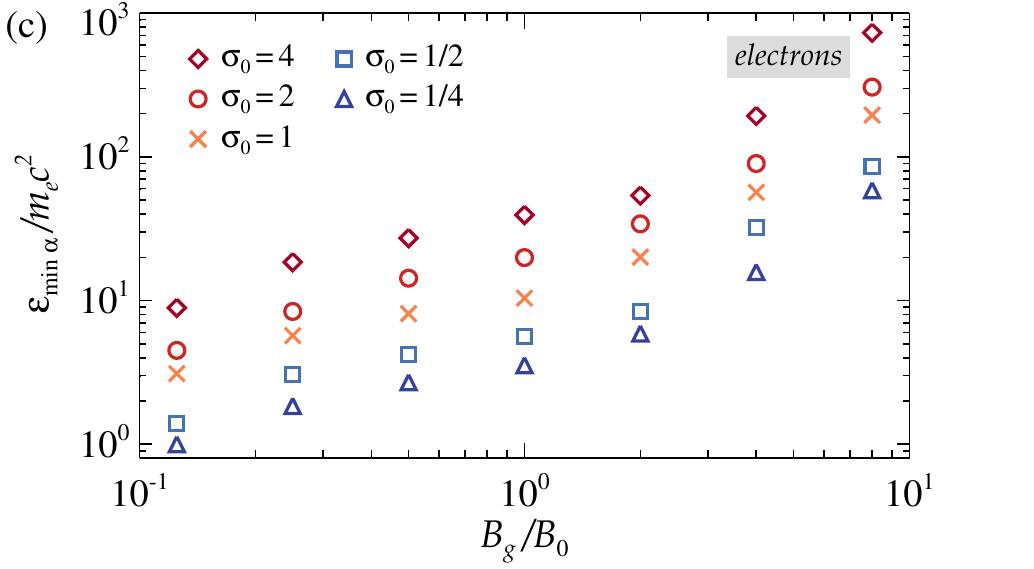}
  \hfill
  \includegraphics[width=8.65cm]{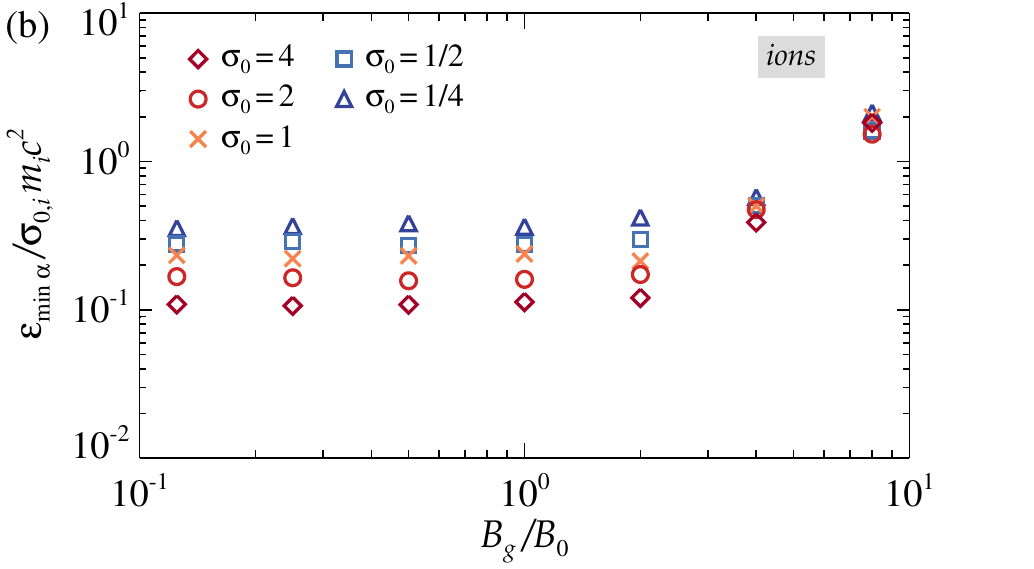}
  \includegraphics[width=8.65cm]{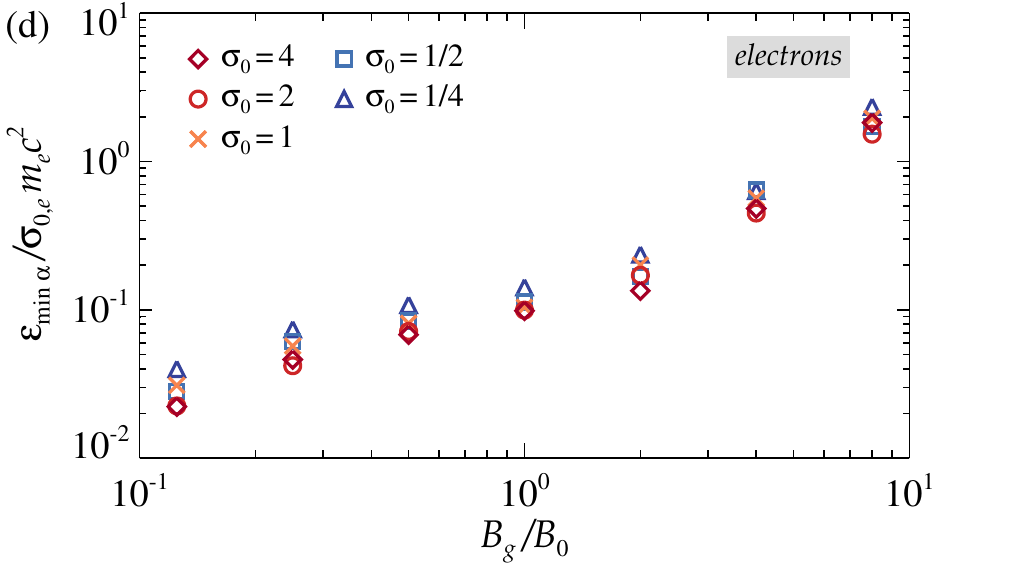}
\caption{Particle kinetic energy at which the pitch angle $\alpha$ reaches its minimum, $\varepsilon_{\min \alpha}$ for ions (left column) and electrons (right column) for the different dimensionless guide field strengths $B_g/B_0$ and various plasma magnetizations $\sigma_0$. Symbols represent different values of $\sigma_0$: $\sigma_0 = 4$ ({\large$\diamond$} in dark red), $\sigma_0 = 2$ ({\large$\circ$} in light red), $\sigma_0 = 1$ ($\times$ in orange), $\sigma_0 = 1/2$ ($\square$ in light blue), and $\sigma_0 = 1/4$ ($\vartriangle$ in dark blue). Normalization of $\varepsilon_{\min \alpha}$ by $m_s c^2$ ($s=i$ for ions, $s=e$ for electrons) in the top row and by $\sigma_{0,i} m_i c^2 = \sigma_{0,e} m_e c^2$ in the bottom row to emphasize the relationship between plasma magnetization and $\varepsilon_{\min \alpha}$.}
\label{fig11}
\end{figure*} 

\begin{figure*}
 \centering 
  \includegraphics[width=8.65cm]{Fig_ions}
  \includegraphics[width=8.65cm]{Fig_electrons}
  \hfill
  \includegraphics[width=8.65cm]{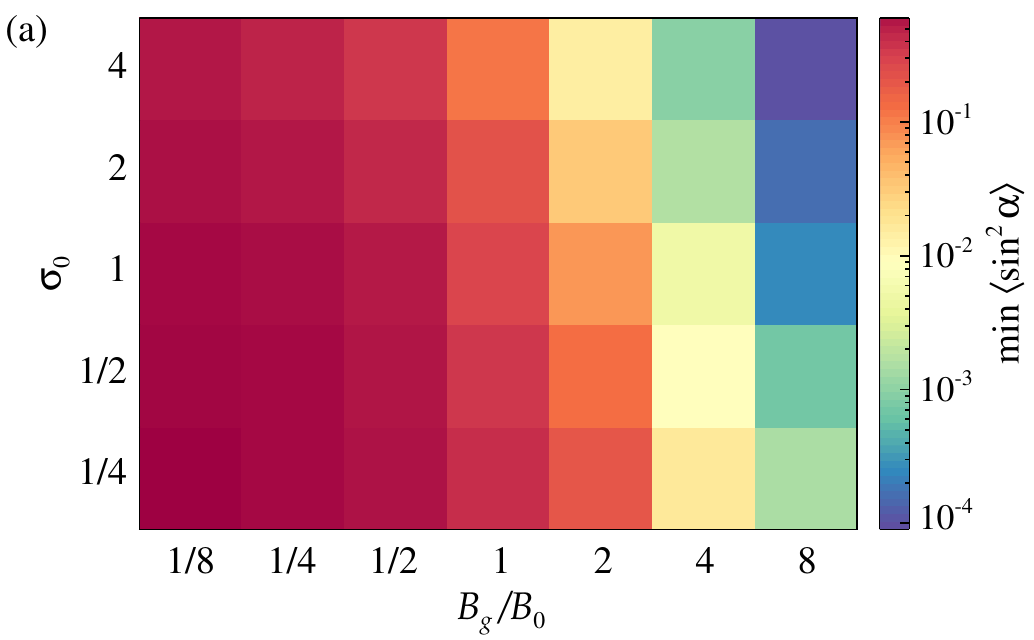}
  \includegraphics[width=8.65cm]{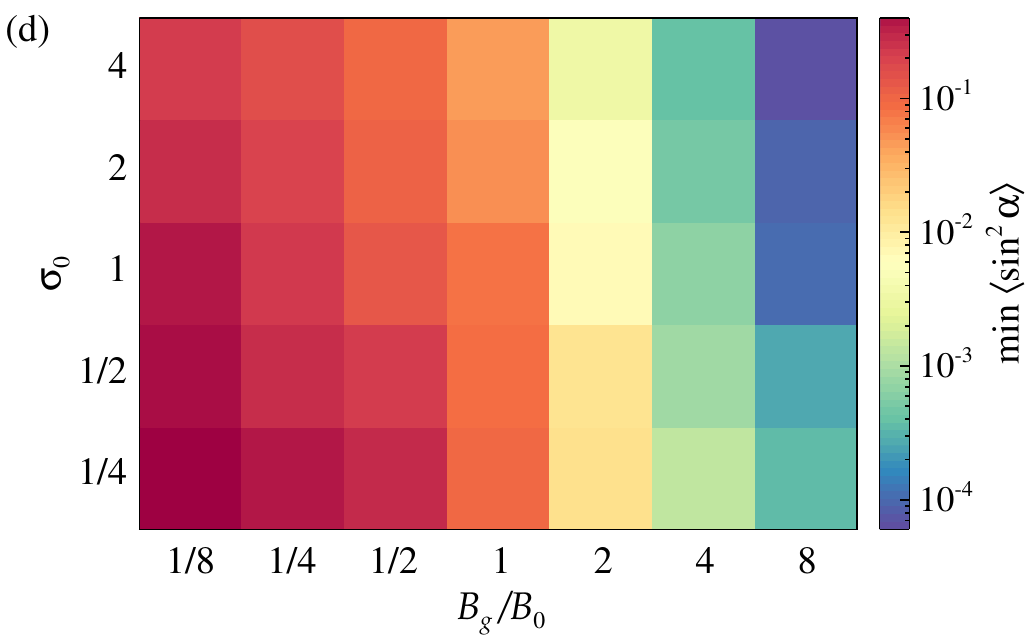} 
  \hfill
  \includegraphics[width=8.65cm]{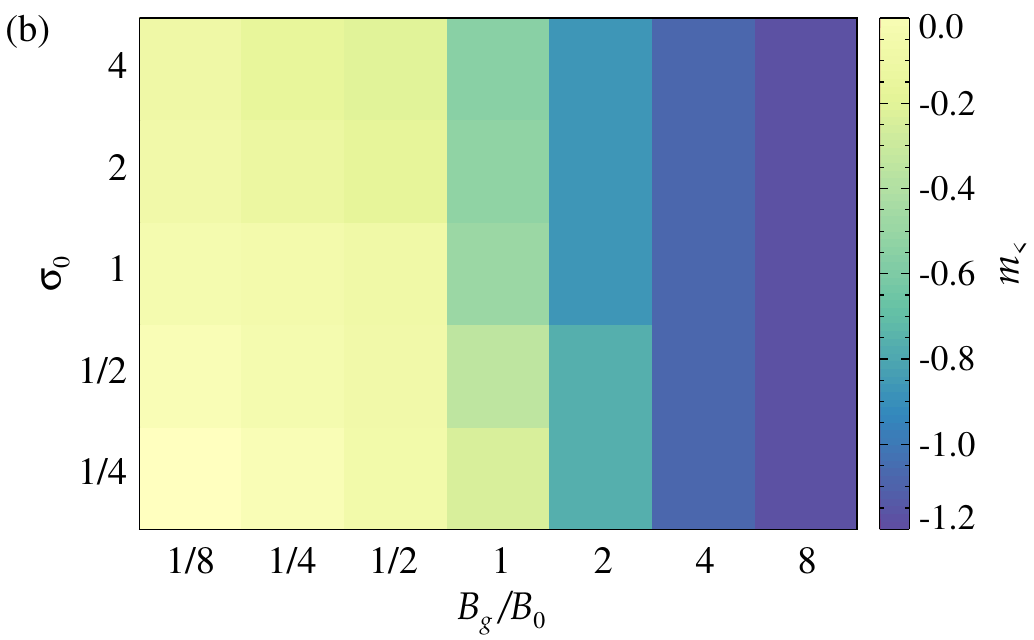}
  \includegraphics[width=8.65cm]{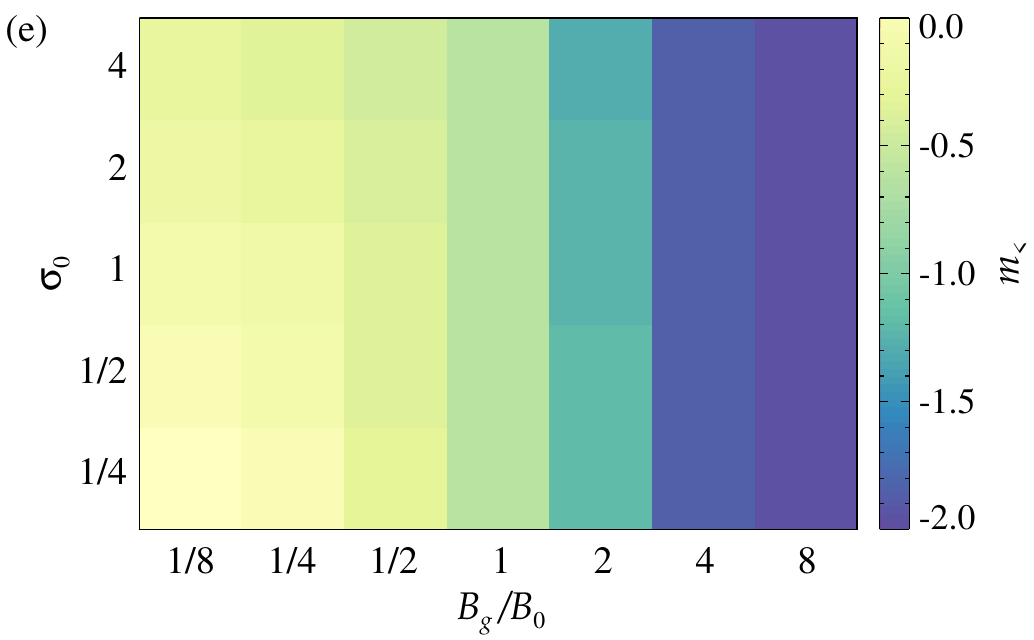}
  \hfill
  \includegraphics[width=8.65cm]{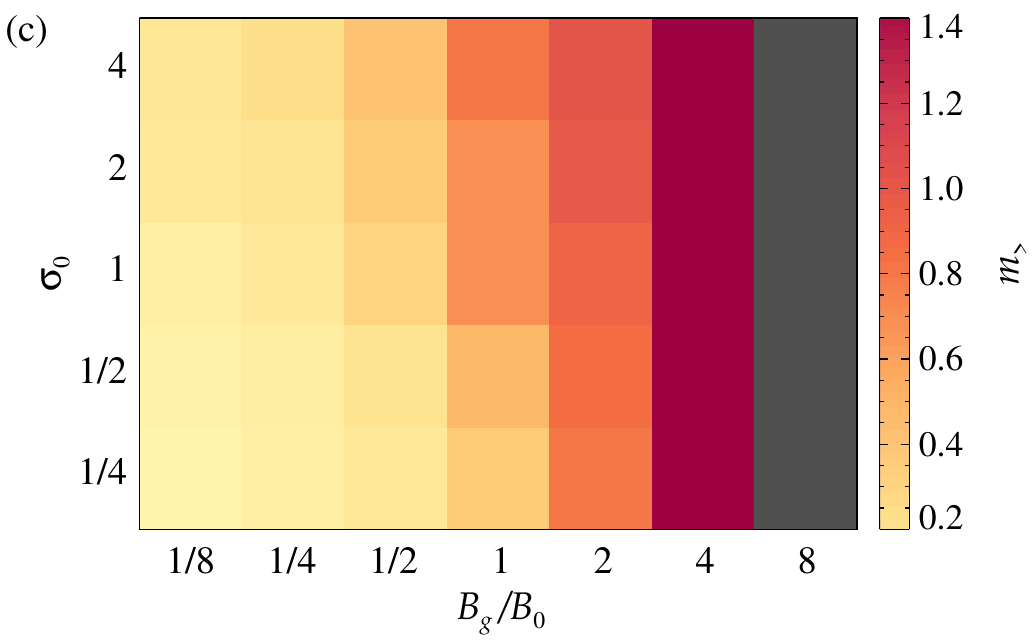}
  \includegraphics[width=8.65cm]{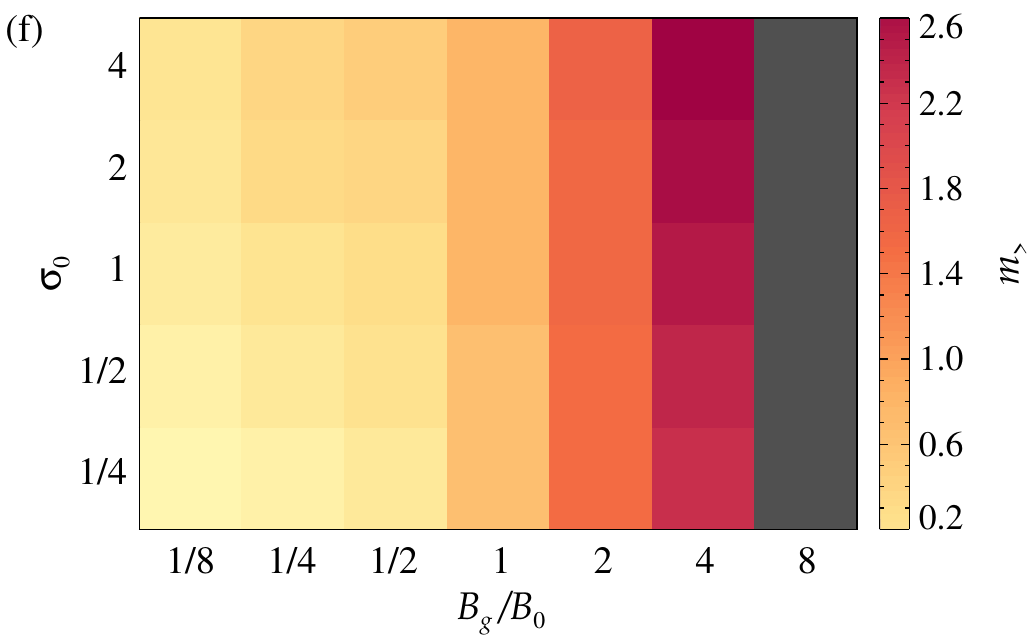}
\caption{2D histograms providing the pitch-angle anisotropy characterization for ions (left) and electrons (right), across a range of dimensionless guide field strengths $B_g/B_0$ and plasma magnetization $\sigma_0$. The top row (panels (a) and (d)) displays the peak anisotropy $\min \langle \sin^2 \alpha \rangle$, the middle row (panels (b) and (e)) shows the power-law indices $m_<$, while the bottom row (panels (c) and (f)) presents the power-law indices $m_>$. The dark gray shading in panels (c) and (f) denotes the absence of a power-law range with slope $m_>$ for $B_g/B_0 = 8$.}
\label{fig12}
\end{figure*} 

We derive the parameters governing the power laws in Eqs. (\ref{eq:dNdgamma}) and (\ref{eq:alphagamma}) directly from our simulation data. In Figure \ref{fig9}, we report the break energy $\varepsilon_0$ for ions (left column) and electrons (right column). The top row displays $\varepsilon_0$ normalized by $m_s c^2$ ($s=i,e$), while the bottom row normalizes it by $\sigma_{0,s} m_s c^2$ in order to highlight the scaling with the magnetization of the upstream particles. From the simulation campaign we obtain
\begin{equation}
   \varepsilon_0 = \kappa_s \sigma_{0,s} m_s c^2  \, ,
\end{equation}
with $\kappa_s \sim 1$ for $s=i$ (see Fig. \ref{fig9}(b)) and $\kappa_s \sim 0.1$ for $s=e$ (see Fig. \ref{fig9}(d)). 
$\kappa_s$ displays minimal variation with respect to $B_g/B_0$. Specifically, for ions, $\kappa_i \simeq 0.5$ for $B_g/B_0 \ll 1$ and $\kappa_i \simeq 1$ for $B_g/B_0 \gtrsim 1$. 
Similarly, for electrons, $\kappa_e \simeq 0.1$ for $B_g/B_0 \ll 1$ and $\kappa_e \simeq 0.15$ for $B_g/B_0 \gtrsim 1$.

In Figure \ref{fig10}, we report the spectral indices obtained for ions (left column) and electrons (right column). The spectral index $p_<$ is ultra-hard ($p_< \lesssim 1$) across all the explored parameter domain except for ions when $\sigma_{0,i} \simeq \sigma_0 \lesssim 1$ and $B_g/B_0 \gtrsim 1$ (see Fig. \ref{fig10}(a)). The ultra-hard slopes observed for electrons in all parameter space (see Fig. \ref{fig10}(c)) align with findings in pair plasma simulations \citep{Comisso23}, where the magnetization of the upstream electrons was much larger than unity (in our simulations, $\sigma_{0,e} = \sigma_0 (1+m_i/m_e) \gg 1$ for all cases). 
In the upper-energy range (Figs. \ref{fig10}(b) and \ref{fig10}(d)), both ions and electrons exhibit power-law indices of $p_> \sim 2$ when $B_g/B_0 \lesssim 1$ and $\sigma_0 \gtrsim 1$ \citep[for electrons, see also][]{Petropoulou18,LiX23, Zhang23ApJL}. 
At lower $\sigma_0$, for ions (Fig. \ref{fig10}(b)) $p_> \sim 3$ for $B_g/B_0 \sim 1/2$ and $\sigma_0 \sim 1$, and $p_> \sim 4$ for $B_g/B_0 \sim 1/2$ and $\sigma_0 \sim 1/4$ \citep[see also][]{Zhang21}. 
For $B_g/B_0 \gtrsim 1$, the upper-energy power law is either indiscernible or absent for the considered values of $\sigma_0$. 
Conversely, electrons exhibit power laws in the high-energy range even for $B_g/B_0 \gtrsim 1$, with $p_> \sim 3-4$ for  $\sigma_0 \sim 1/4 - 4$ \citep[see also][]{LiX23}.

In Figure \ref{fig11}, we present the measurements of $\varepsilon_{\min \alpha}$ for ions (left column) and electrons (right column). In the top row, $\varepsilon_{\min \alpha}$ is normalized by $m_s c^2$ ($s=i,e$), while in the bottom row, it is normalized by $\sigma_{0,s} m_s c^2$.
The value of $\varepsilon_{\min \alpha}$ is sensitive to the strength of the guide field. 
When $B_g/B_0 \ll 1$, pitch-angle anisotropy is weak for electrons and essentially absent for ions. In this regime, $\varepsilon_{\min \alpha}$ resides at low energies, below the break energy $\varepsilon_0$. 
As $B_g/B_0$ approaches $1$, pitch-angle anisotropy becomes more pronounced, especially for electrons, in which case $\varepsilon_{\min \alpha} \sim \varepsilon_0$ \citep{Comisso23}. In this regime, ion pitch-angle anisotropy remains weak, and $\varepsilon_{\min \alpha}$ scales sublinearly with $\sigma_{0,s}$. 
Conversely, for $B_g/B_0 \gg 1$, $\varepsilon_{\min \alpha}$ scales with $\sigma_{0,s}$ for both ions and electrons. For the cases of stronger guide field strength considered here ($B_g =8 B_0$) the maximum anisotropy reaches $\varepsilon_{\min \alpha} \simeq 2 \sigma_{0,s} m_s c^2$ (see Figs. \ref{fig11}(b) and \ref{fig11}(d)).
This energy scale also corresponds to the energy cutoff $\varepsilon_{\rm cut}$ when $B_g/B_0 \gg 1$. On the other hand, in the opposite limit $B_g/B_0 \ll 1$, both ions and electrons are further energized to higher energies up to the Hillas limit \citep{Hillas84} of the reconnection layer. This occurs when the particle gyroradius matches the width of the reconnection layer, namely $r_g(\varepsilon_{\rm cut}) \simeq \langle R_{\rm rec} \rangle 2 L_x$ \citep[see also][]{ZhangH21,LiX23}, implying $\varepsilon_{\rm cut} \simeq 2 \langle R_{\rm rec} \rangle \sqrt{\sigma_0} L_x/d_e$ for ultrarelativistic particles.

In Figure \ref{fig12}, we report the minimum of $\langle \sin^2 \alpha \rangle$ (top row), along with the power-law slopes $m_<$ (middle row) and $m_>$ (bottom row), obtained for ions (left column) and electrons (right column). 
$\min \langle \sin^2 \alpha \rangle$ decreases with higher values of $B_g/B_0$ and $\sigma_0$, as evidenced by the decreasing trend observed from the bottom-left to the upper-right corners in Figs. \ref{fig12}(a) and \ref{fig12}(d). For the strongest guide field and higher magnetization explored here, $\langle \sin^2 \alpha \rangle$ drops four orders of magnitude below the isotropic expectation of $2/3$.

A transition from moderate to very strong depletion of $\langle \sin^2 \alpha \rangle$ occurs around $B_g/B_0 \sim 1$ for both ions and electrons. When $B_g/B_0 \ll 1$, $\langle \sin^2 \alpha \rangle$ is close to isotropy and the slopes $m_<$ and $m_>$ are small. At $B_g/B_0 \sim 1$, typical values for $|m_<|$ and $|m_>|$ are around $1$, slightly lower for ions compared to electrons. 
As $B_g/B_0$ increases, eventually the positive power-law slope $m_>$ disappears (as observed in Figs. \ref{fig12}(c) and \ref{fig12}(f) for $B_g = 8 B_0$) since at excessively large $B_g/B_0$, the accelerated particles are unable to effectively undergo pitch angle isotropization. 
In the asymptotic limit $B_g/B_0 \gg 1$, the influence of pitch-angle scattering on pitch-angle anisotropy becomes negligible, as $t_{\rm scatt} \sim (1 + B_g^2/B_0^2)L_x/v_{A0}$ \citep{Comisso23}.
For particles with a Lorentz factor $\gamma_{\rm th} \sim 1$, assuming that all energy gain occurs in the magnetic field-aligned direction via $E_\parallel$, then $v_\parallel/c \sim (1 - 1/\gamma)^{1/2}$ for $\Delta \gamma = \gamma - \gamma_{\rm th} \gg 1$. 
Therefore, for $B_g/B_0 \gg 1$ and $\Delta \varepsilon/m_s c^2 \gg 1$, the pitch-angle anisotropy scales as $\langle \sin^2 \alpha \rangle \propto \varepsilon^{-2}$ for $1 \ll \varepsilon < \varepsilon_{\rm cut}$. 
In our simulations, the condition $\Delta \varepsilon/m_e c^2 \gg 1$ is met for electrons, for which we find $m_< \simeq -2$ for $B_g = 8 B_0$ (Fig. \ref{fig12}(e)).

\section{Astrophysical Implications}\label{sec:discussion} 

The pitch-angle anisotropy of reconnection-accelerated particles plays a central role in determining the properties of their emitted radiation \citep[e.g.][]{DermerMenon2009}. 
In the comoving plasma frame, the instantaneous particle energy loss rate due to synchrotron radiation for particles with mass $m_s$ and charge $Ze$ is given by
\begin{equation}
- m_s c^2 {\left( {\frac{d \gamma}{dt}} \right)}_{\rm{sync}} = 2 \sigma_T Z^4 {\left( {\frac{m_e}{m_s}} \right)}^2 c U_B  {\left( {\frac{v}{c}} \right)}^{2} \gamma^2 \sin^2 \alpha \, .
\end{equation}
Here, $U_B$ indicates the magnetic energy density and $\sigma_T$ denotes the Thomson cross section for an electron. 
Since $- {\left( {{d \gamma}/{dt}} \right)}_{\rm{sync}} \propto \sin^2 \alpha$, and strong deviations from pitch-angle isotropy can arise when $B_g/B_0 \gtrsim 1$ (see Figs. \ref{fig12}(a) and \ref{fig12}(d)), the pitch-angle anisotropy of reconnection-accelerated particles can drastically suppress synchrotron luminosity compared to the isotropic case. This synchrotron suppression opens up the possibility for other emission mechanisms (e.g., inverse Compton scattering) to prevail even in regions characterized by strong magnetic fields where synchrotron emission would be presumed dominant under isotropic pitch angle assumptions (see \citet{Sobacchi2021MNRAS,Sobacchi_SSC2021}).

The pitch angle distribution of the accelerated particles is not only anisotropic but energy-dependent, $\langle \sin^2 \alpha \rangle \propto \varepsilon^m$ (see Figs. \ref{2D_PA}-\ref{fig6}). This energy dependence plays a significant role in shaping the synchrotron spectrum produced by the accelerated particles. 
For a power-law energy distribution of relativistic electrons, $N(\varepsilon) d\varepsilon \propto \varepsilon^{-p} d\varepsilon$, under the usual assumption of pitch-angle isotropy, $\langle \sin^2 \alpha \rangle = 2/3$, the radiation power spectrum follows the standard expression $F_\nu(\nu) \propto\nu^{(1-p)/2}$  \citep{RybLig79}, which relates the particle spectral index $p$ with the radiation spectral index for optically thin synchrotron radiation.
However, the energy-dependent pitch angle anisotropy leads to a different radiation power spectrum, given by \citep{Comisso20ApJL,Tavecchio2020MNRAS,Comisso23}  
\begin{equation}
\label{eq:Fnu}
F_\nu(\nu) \propto \nu^{(2-2p+m)/(4+m)} \, .
\end{equation}
Equation (\ref{eq:Fnu}) indicates that an anisotropic pitch angle distribution has a substantial effect on the synchrotron spectrum, causing hardening for $m>0$ and softening for $m<0$ compared to the isotropic assumption.
For scenarios where $B_g/B_0 \sim 1$ and $\sigma_0 \gtrsim 1$, using characteristic values of $p_< \sim 0.4$ and $m_< \sim -0.7$ for the lower-energy slopes, and $p_> \sim 2$ and $m_> \sim 1$ for the higher-energy slopes (refer to Fig. \ref{fig5}(b)), we deduce a radiation power spectrum of $F_\nu(\nu) \propto\nu^{0.15}$ in the $m_<$ range, and $F_\nu(\nu) \propto\nu^{-1/5}$ in the $m_>$ range (contrasting with the anticipated $F_\nu(\nu) \propto\nu^{0.3}$ and $F_\nu(\nu) \propto\nu^{-1/2}$ spectra expected under the assumption of energy-independent pitch angles). 
In general, for $B_g/B_0 \gtrsim1$ and $\sigma_0 \gtrsim 1$, typical $p_>$ and $m_>$ values (refer to Figs. \ref{fig10} and \ref{fig12}) result in a synchrotron spectrum that is essentially flat, $F_\nu(\nu) \propto\nu^{0}$. 
Note that the synchrotron spectrum in the $m_>$ range spans a broader range of frequencies compared to the isotropic model (the opposite holds for the $m_<$ range). 
Specifically, this occurs across a frequency span determined by the ratio ${\nu_c(\gamma_{\rm iso})}/{\nu_c(\gamma_{\min \alpha})} \simeq \left({\gamma_{\rm iso}}/{\gamma_{\min \alpha}} \right)^{(4+m_>)/2}$. 
For $B_g/B_0 \sim 1$ and $\sigma_0 \gtrsim 1$, we find $\gamma_{\rm iso}/\gamma_{\min \alpha} \sim 40$ (see Fig. \ref{fig5}(b)). Consequently, for $m_> \sim 1$, the synchrotron spectrum's hardening due to pitch-angle anisotropy is expected to span a frequency range of $4$ decades.

As particles emit synchrotron radiation, they lose energy over a characteristic timescale given by $t_{\rm{cool}} = -\gamma/({d \gamma}/{dt}) = {m_s^3 c}/({2 \sigma_T m_e^2 Z^4 U_B \gamma \sin^2 \alpha})$. Particles are strongly cooled if they radiate a significant fraction of their energy in a timescale shorter than $L_x/\langle v_{\rm out} \rangle$. Using Eq. (\ref{eq:vout}), the Lorentz factor of a particle that cools in such a timescale can be expressed as  
\begin{equation} \label{gamma_cool}
\gamma_{\rm{cool}} = \frac{m_s c^2}{2 \sigma_T Z^4 U_B L_x} \frac{1}{\sin^2 \alpha} \left( {\frac{m_s}{m_e}} \right)^{2} \left( {\frac{\sigma_0}{1 + \sigma_0 + \sigma_g}} \right)^{1/2}\, ,
\end{equation}
Particles with $\gamma > \gamma_{\rm{cool}}$ are in a fast cooling regime, while particles with $\gamma < \gamma_{\rm{cool}}$ are in a slow cooling regime. 
The particle spectrum is affected by cooling on the timescale $L_x/\langle v_{\rm out} \rangle$ if it extends beyond $\gamma_{\rm{cool}}$. 
Given that $\gamma_{\rm cool} \propto 1/\sin^2 \alpha$, when $B_g/B_0 \gtrsim 1$ cooling becomes significantly less severe than what would be expected under the assumption of isotropic pitch angles. 
Assuming $\sin^2 \alpha \propto \gamma^m$, $t_{\rm cool} \propto \gamma^{-(1+m)}$. This implies that within the $m_<$ range, synchrotron cooling becomes stronger for lower-energy particles when $m_< < -1$. This results in a hardening of the particle energy spectrum within the range governed by the $m_<$ slope \citep{Comisso21}. The faster cooling of lower-energy particles ceases when $m_< \geq -1$. This suggests that strong synchrotron cooling tends to induce and sustain a pitch-angle anisotropy with $m_< \sim -1$. For $B_g/B_0 \gtrsim1$, employing $p_< \sim 0.5$ (see Fig. \ref{fig10}(c)) with $m_< \sim -1$, we deduce from Eq. (\ref{eq:Fnu}) a flat synchrotron spectrum, $F_\nu(\nu) \propto \nu^{0}$.

Pitch-angle anisotropy allows particles to emit synchrotron radiation beyond the ideal MHD limit of $h \nu_{\rm burnoff} \simeq 160 \, {\rm{MeV}}$ in the plasma comoving frame \citep{Guilbert83,deJager96}. The radiation reaction limit is dictated by the balance between the accelerating electric force on the particle and the radiation reaction force due to synchrotron emission. Taking $F_{RR}^{\rm{sync}}  =  2 \sigma_T Z^4 (m_e/m_s)^2 U_B \gamma^2 \sin^2 \alpha$ for $\gamma \gg 1$, the radiation-reaction-limited Lorentz factor is given by 
\begin{equation} \label{}
\gamma_{\rm{rad}} =  \frac{m_s}{m_e} \left(\frac{e E_{\rm rec}}{2 \sigma_T Z^3 U_B \sin^2 \alpha}\right)^{1/2} 
\end{equation}
in the comoving frame. For protons, $\gamma_{\rm{rad}}$ is hardly a constraint given the mass ratio $m_p/m_e \simeq 1836$. For electrons, the critical synchrotron photon energy emitted by a population of particles limited by radiation reaction is given by
\begin{equation} \label{eq:hnu_max}
h \nu_{\rm c, max} = \frac{3}{2} \hbar \gamma_{\rm{rad}}^2 \omega_L {\langle \sin \alpha \rangle} = \frac{9}{4 \alpha_{\rm fs}} \frac{E_{\rm rec} m_e c^2}{\langle \sin \alpha \rangle (B_0^2 + B_g^2)^{1/2}} \, ,
\end{equation}
where $\hbar$ is the reduced Planck constant, and $\alpha_{\rm fs} \simeq 1/137$ is the fine-structure constant. With the aid of Eq. (\ref{eq:Rrec}), we can simplify Eq. (\ref{eq:hnu_max}) in the relevant limits
\begin{equation}
h \nu_{\rm c, max} \simeq 160 \frac{\epsilon_{\rm rec}}{\langle \sin \alpha \rangle} \frac{v_{A0}}{c} \, {\rm{MeV}} \, , \qquad \frac{B_g}{B_0} \ll \frac{c}{v_{A0}} \, ,
\end{equation}
\begin{equation}
h \nu_{\rm c, max} \simeq 160 \frac{\epsilon_{\rm rec}}{\langle \sin \alpha \rangle} {\left( {\frac{B_0}{B_g}} \right)}^2 {\rm{MeV}} \, , \qquad \frac{B_g}{B_0} \gg \frac{c}{v_{A0}} \, .
\end{equation}
For ${B_g}/{B_0} \ll {c}/{v_{A0}}$, strong deviations from pitch angle isotropy are not expected. However, the ideal MHD limit of $160 \, {\rm{MeV}}$ can be bypassed, since $E_{\rm rec}$ can accelerate particles in regions where $|\bm{B}| \simeq B_g \ll B_0$ \citep{Cerutti13,Yuan16,Hayk19,Chernoglazov23}.
Conversely, for ${B_g}/{B_0} \gg {c}/{v_{A0}}$, a significant magnetic field is always present throughout the reconnection layer. Considering the high magnetization limit $\sigma_0 \gg 1$, for which $c/v_{A0} \simeq 1$, electrons emitting at $\gamma = \sigma_{0,e}$ produce radiation above $160 \, {\rm{MeV}}$ if the electron magnetization satisfies $\sigma_{0,e} > \epsilon_{\rm rec}^{-1} (B_g/B_0)^2$, where we used $\langle \sin \alpha \rangle \simeq 1/\sigma_{0,e}$. 
Therefore, in the presence of a strong guide field, pitch angle anisotropy enables electrons to emit radiation above the ideal MHD limit of $160 \, {\rm{MeV}}$.

Another important aspect of the pitch-angle anisotropy is its impact on the degree of linear polarization of synchrotron radiation. 
For a power-law distribution of relativistic electrons with an energy-independent pitch angle, the degree of linear polarization is linked to the power law index of the energy distribution of the emitting particles by the standard formula $\Pi_{\rm lin} = (p+1)/(p+7/3)$ \citep{RybLig79}, applicable for a uniform magnetic field. However, when $\langle \sin^2 \alpha \rangle \propto \varepsilon^m$, this relation for the degree of linear polarization modifies to 
\citep{Comisso23}
\begin{equation} \label{Com_linpolar}
\Pi_{\rm lin} = \frac{p+1}{p+7/3+m/3} \, .
\end{equation} 
Therefore, pitch angle anisotropy reduces the degree of linear polarization with respect to the isotropic case for $m >0$, while it increases it for $m < 0$. In scenarios characterized by strong ordered magnetic fields, with $B_g/B_0 \gg 1$, where a high degree of linear polarization is expected, we obtained an extended range $m = m_<$, with $m_< \sim -2$ for electrons (see Fig. \ref{fig12}(e)). Using Eq. (\ref{Com_linpolar}) with  $p_< \sim 1$, we deduce a linear polarization degree of $\Pi_{\rm lin} \simeq 75\%$, higher than the value of $\Pi_{\rm lin} \simeq 60\%$ that would have been inferred without accounting for the pitch angle anisotropy. On the other hand, for reconnecting current sheets that are embedded in a turbulent environment, the disordered magnetic field fluctuations would significantly reduce the overall polarization degree.
Additionally, it's worth noting that pitch angle anisotropy can further produce a significant degree of circular polarization of the emitted radiation \citep[see][]{Comisso23}, characterized by $\Pi_{\rm cir} \gg 1/\gamma_e$, where $\gamma_e$ is the Lorentz factor of the radiating electrons.

The energy-dependent pitch-angle anisotropy can also lead to highly beamed radiation aligned with the magnetic field direction. 
Particles with $\gamma \sim \gamma_{\min \alpha}$ are expected to predominantly emit radiation in the direction of $B_0$ when $B_g/B_0 \ll 1$ \citep{Cerutti2012ApJL} and in the direction of $B_g$ when $B_g/B_0 \gtrsim 1$ \citep{Comisso20ApJL}. 
When radiation emission occurs on a timescale longer than that required for imprinting the pitch-angle anisotropy but shorter than the scattering timescale, i.e. $t_{\min \alpha} \ll t_{\rm cool} \ll t_{\rm scatt}$, the emitted radiation, whether via synchrotron or other mechanisms (e.g., inverse Compton scattering), is beamed into a solid angle $\Delta \Omega \sim \langle \sin^2 \alpha \rangle$ around the local magnetic field. The extent of $\Delta \Omega$ is tied to the particle's energy due to the energy-dependent nature of the pitch-angle anisotropy, as outlined in Eq. (\ref{eq:alphagamma}). 
In the presence of strong beaming, the emission from a reconnecting current sheet is observable only if the local magnetic field is closely aligned with the observer's line of sight, which can lead to rapid variability of the emission \citep{Sob23}.
When multiple active radiation beams are present and the beams are distributed isotropically, the variability timescale $\delta T$ for an emission event spanning a timescale $T$ depends on the pitch angle as
\begin{equation} 
\frac{\delta T}{T} \sim N_{\alpha} \langle \sin^2 \alpha \rangle \, ,
\end{equation}
where $N_{\alpha}$ represents the number of active beams. In this scenario, the variability resulting from pitch-angle anisotropy is frequency-dependent, in contrast to the achromatic variability associated with Doppler beaming from bulk motions \citep{Lyutikov06,NarayanKumar09,Lazar09,Gia09}. 
One might anticipate that the variability resulting from pitch-angle anisotropy is more prevalent than that resulting from Doppler beaming associated with bulk motions in the reconnection layer. From Eq. (\ref{eq:vout}), the maximum Lorentz factor of the bulk flow within the reconnection layer is given by  
$\Gamma_{\rm{bulk, max}} \simeq   \left[{(1+\sigma_g+\sigma_0)}/{(1+\sigma_g)}\right]^{1/2}$.
This implies that even a moderate guide field strength $B_g \gtrsim B_0/3$ or an overall magnetization $\sigma_0 \lesssim 1$ results in $\Gamma_{\rm{bulk, max}} \sim 1$, effectively eliminating the possibility of bulk outflow beaming. On the other hand, the beaming associated with pitch-angle anisotropy only requires high electron magnetization, $\sigma_{0,e} \gg 1$.

The pitch angle anisotropy induced by magnetic reconnection also leads to temperature distributions that are anisotropic. In the regime where $B_g/B_0 \gg 1$, the pitch angle anisotropy is most pronounced, with the majority of the magnetic energy released by magnetic reconnection being channeled into electrons (see Fig. \ref{fig7}). As a result, in this regime $T_{e,\parallel} \gg T_{e,\perp}$, where $T_{e,\parallel} = P_{e,\parallel}/k_B n_e$ and $T_{e,\perp} = P_{e,\perp}/k_B n_e$ represent the parallel and perpendicular temperatures, respectively, with $P_{e,\parallel}$ and $P_{e,\perp}$ indicating the electron pressures parallel and perpendicular to the magnetic field direction. 
This pressure anisotropy persists against the firehose instability provided that $1 -P_{e,\perp}/P_{e,\parallel} - 2/\beta_{e,\parallel} < 0$ \citep{Gary93}.
Therefore, under highly magnetized conditions ($\beta_{e,\parallel} < 1$), the temperature anisotropy established by magnetic reconnection persists, with $T_{e,\parallel} \gg T_{e,\perp}$ for $B_g/B_0 \gg 1$. 
This anisotropy has significant implications for the interpretation of synchrotron radiation from astrophysical plasmas, as the majority of models assume isotropic particle distributions, thereby substantially underestimating the parallel temperature of electrons.

\section{Conclusions}

We investigated the simultaneous generation of energetic particles and pitch-angle anisotropy resulting from magnetic reconnection in ion-electron plasmas. 
With the aid of first-principles PIC simulations, we demonstrated that particle acceleration driven by magnetic reconnection produces anisotropic distributions $f_s \left( {|\cos \alpha|,\varepsilon} \right)$ for both ions and electrons. 
In general, the process of particle acceleration through reconnection results in broken power laws in $f_s (\varepsilon)$ and the mean of $\sin^2 \alpha$ at a given $\varepsilon$. 
We show that the particle energy spectra and pitch-angle anisotropy are controlled by the relative strength of the guide field compared to the reconnecting magnetic field, $B_g/B_0$, as well as the plasma magnetization $\sigma_0$. 

The break energy that separates the two power-law ranges in the particle energy spectra, $\varepsilon_0$, scales linearly with the magnetization parameter and displays minimal sensitivity to the strength of the guide field. This break energy represents a considerable fraction of the mean energy per particle after magnetic field dissipation, $\varepsilon_0 = \kappa_s \sigma_{0,s} m_s c^2$, with $\kappa_i \simeq 0.5-1$ for ions and $\kappa_e \simeq 0.1-0.15$ for electrons. 
Below $\varepsilon_0$, the particle energy spectrum is governed by particle injection. The injection slope exhibits remarkable hardness ($p_< \lesssim 1$) with minimal sensitivity to the guide field strength when $\sigma_0 \gg 1$. In contrast, the power-law range above $\varepsilon_0$ displays considerable sensitivity to the strength of the guide field. For both ions and electrons, the spectral index $p_>$ typically falls within the range $p_> \sim [2,4]$ across the explored parameter space.
The extension of this second power-law range depends on the guide field strength, with $B_g/B_0 \gg 1$ resulting in a cutoff energy $\varepsilon_{\rm cut} \sim 2 \sigma_{0,s} m_s c^2$, while for $B_g/B_0 \ll 1$, the cutoff energy approaches the Hillas limit of the reconnection layer. 
The ratio $B_g/B_0$ also regulates the redistribution of magnetic energy between ions and electrons, with $\Delta{E_i} \gg \Delta{E_e}$ for $B_g/B_0 \ll 1$, $\Delta{E_i} \sim \Delta{E_e}$ for $B_g/B_0 \sim 1$, and $\Delta{E_i} \ll \Delta{E_e}$ for $B_g/B_0 \gg 1$, with $\Delta{E_i}/\Delta{E_e}$ approaching unity when $\sigma_0 \gg 1$. 

The mean pitch angle at a given energy reaches the strongest deviation from isotropy at a characteristic energy $\varepsilon_{\min \alpha}$.
For $B_g/B_0 \ll 1$, deviations of $\langle \sin^2 \alpha \rangle$ from isotropy are modest, and $\varepsilon_{\min \alpha} \ll \varepsilon_0$ for both ions and electrons. 
Conversely, for $B_g/B_0 \sim 1$, $\langle \sin^2 \alpha \rangle$ departs significantly from isotropy, with $\varepsilon_{\min \alpha} \sim \varepsilon_0$ for electrons, while ions' $\varepsilon_{\min \alpha}$ scales sublinearly with $\sigma_{0}$ within the range of values considered in this study. For values of $\sigma_{0}$ that fall well into the asymptotic limit $\sigma_{0} \gg 1$, the behavior of the electron-ion plasma approaches that of a pair plasma. In such cases, ions' $\varepsilon_{\min \alpha}$ mirrors that of electrons' $\varepsilon_{\min \alpha}$ \citep{Comisso23}. 
For $B_g/B_0 \gg 1$, deviations of $\langle \sin^2 \alpha \rangle$ from isotropy increase continuously with particle energy, and $\varepsilon_{\min \alpha} \sim \varepsilon_{\rm cut} \sim 2 \sigma_{0,s} m_s c^2$ for both ions and electrons. 
Below $\varepsilon_{\min \alpha}$, $\langle \sin^2 \alpha \rangle \propto \varepsilon^{m_<}$, with negative slopes $m_<$ decreasing with increasing $B_g/B_0$ and $\sigma_0$. 
Above $\varepsilon_{\min \alpha}$, $\langle \sin^2 \alpha \rangle \propto \varepsilon^{m_>}$, with positive slopes $m_>$ increasing with $B_g/B_0$ and $\sigma_0$.
The anisotropy is primarily governed by the ratio of the guide field's strength to that of the reconnecting field, and when $B_g/B_0$ falls well into the asymptotic limit $B_g/B_0 \gg 1$, the rising range with a positive slope eventually disappears. 
In the limit for which $B_g/B_0 \gg 1$ and $\Delta \varepsilon/m_s c^2 \gg 1$, the pitch angle anisotropy reaches its strongest level and scales as $\langle \sin^2 \alpha \rangle \propto \varepsilon^{-2}$ for $1 \ll \varepsilon < \varepsilon_{\rm cut}$. 

The anisotropic particle distributions that arise from magnetic reconnection carry significant astrophysical implications. 
These anisotropic distributions lead to a radiation power spectrum $F_\nu(\nu) \propto \nu^{(2-2p+m)/(4+m)}$, resulting in hardening in the $m=m_>$ range and softening in the $m=m_<$ range compared to isotropic expectations.
The degree of linear polarization is also affected, with an increase in the linear polarization degree in the $m=m_<$ range and a reduction in the $m=m_>$ range, as indicated by the relation $\Pi_{\rm lin} = (p+1)/(p+7/3+m/3)$.
Clearly, the imprinting of small pitch angles on the particles accelerated by magnetic reconnection impacts synchrotron luminosity, progressively reducing it with increasing guide field. 
Pitch-angle anisotropy also imposes stricter conditions for entering the fast cooling regime while enabling synchrotron radiation emission at higher frequencies, especially near the radiation-reaction limit. 
Energy-dependent pitch-angle anisotropy can produce highly beamed radiation aligned with the magnetic field direction, inducing frequency-dependent variability ${\delta T}/{T} \sim N_{\alpha} \langle \sin^2 \alpha \rangle$. 
Moreover, pitch-angle anisotropy leads to anisotropic electron temperatures, with $T_{e,\parallel} \gg T_{e,\perp}$ for $B_g/B_0 \gg 1$. 
All these effects highlight the necessity of incorporating pitch-angle anisotropy in astrophysical models to improve their predictive power and our understanding of astrophysical sources of radiation.

\section*{Acknowledgments}
We acknowledge fruitful discussions with Brian Jiang, Emanuele Sobacchi and Lorenzo Sironi. This research is supported by the NASA ATP award 80NSSC22K0667. We acknowledge computing resources from Columbia University's Shared Research Computing Facility project, which is supported by NIH Research Facility Improvement Grant 1G20RR030893-01, and associated funds from the New York State Empire State Development, Division of Science Technology and Innovation (NYSTAR) contract C090171.

\bibliography{rec_pitchangle_ei}{}
\bibliographystyle{aasjournal}

\end{document}